\documentclass[a4paper,12pt]{article}      
\usepackage{amsmath,amssymb,amsfonts} 
\usepackage{setspace,subfigure}
\usepackage{graphics}                 
\usepackage{color}                    
\usepackage{subfig}
\usepackage{rotating}
\setkeys{Gin}{draft=false}

\title{A Statistical Mechanical Approach for the Computation of the Climatic Response to General Forcings}
\author{Valerio Lucarini [\texttt{v.lucarini@reading.ac.uk}] \\ Department of Meteorology\\ Department of Mathematics\\ Walker Institute for Climate System Research\\ University of Reading, Reading, UK \and Stefania Sarno\\ Walker Institute for Climate System Research\\ University of Reading, Reading, UK }

\date{\today}
\begin{document}
\maketitle
\newpage
\begin{abstract}
\singlespacing
The climate belongs to the class of non-equilibrium forced and dissipative systems, for which most results of quasi-equilibrium statistical mechanics, including the fluctuation-dissipation theorem, do not apply. In this paper we show for the first time how the Ruelle linear response theory, developed for studying rigorously the impact of perturbations on general observables of non-equilibrium statistical mechanical systems, can be applied with great success to analyze the climatic response to general forcings. The crucial value of the Ruelle theory lies in the fact that it allows to compute the response of the system in terms of expectation values of explicit and computable functions of the phase space averaged over the invariant measure of the unperturbed state. We choose as test bed a classical version of the Lorenz 96 model, which, in spite of its simplicity, has a well-recognized prototypical value as it is a spatially extended one-dimensional model and presents the basic ingredients, such as dissipation, advection and the presence of an external forcing, of the actual atmosphere. We recapitulate the main aspects of the general response theory and propose some new general results. We then analyze the frequency dependence of the response of both local and global observables to perturbations having localized as well as global spatial patterns. We derive analytically several properties of the corresponding susceptibilities, such as asymptotic behavior, validity of Kramers-Kronig relations, and sum rules, whose main ingredient is the causality principle. We show that all the coefficients of the leading asymptotic expansions as well as the integral constraints can be written as linear function of parameters that describe the unperturbed properties of the system, such as its average energy. Some newly obtained empirical closure equations for such parameters allow to define such properties as an explicit function of the unperturbed forcing parameter alone for a general class of chaotic Lorenz 96 models. We then verify the theoretical predictions from the outputs of the simulations up to a high degree of precision. The theory is used to explain differences in the response of local and global observables, in defining the intensive properties of the system, which do not depend on the spatial resolution of the Lorenz 96 model, and in generalizing the concept of climate sensitivity to all time scales. We also show how to reconstruct the linear Green function, which maps  perturbations of general time patterns into changes in the expectation value of the considered observable for finite as well as infinite time. Finally, we propose a simple yet general methodology to study general Climate Change problems on virtually any time scale by resorting to only few, well selected simulations, and by taking full advantage of ensemble methods. The specific case of globally averaged surface temperature response to a general pattern of change of the $CO_2$ concentration is discussed. We believe that the proposed approach may constitute a mathematically rigorous and practically very effective way to approach the problem of climate sensitivity and climate change from a radically new perspective.
\end{abstract}
\newpage
\tableofcontents
\newpage
\section{Introduction}

A crucial goal in the study of general dynamical and statistical mechanical systems is to understand how their statistical properties are altered when we introduce a perturbation related to changes in the external forcing or in the value of some internal parameters. The ability to compute the response of the system is of great relevance for purely mathematical reasons as well as in many fields of science and technology.

The climate system is an outstanding example of a non-equilibrium, forced and dissipative complex system, forced in first instance by spatial differences and temporal variability in the net energy flux at the top of the atmosphere. On a macroscopic level, as a result of being far from equilibrium, the climate system behaves as an engine, driven by the temperature difference between a warm and a cold thermal pool, so that the atmospheric and oceanic motions are at the same time the result of the mechanical work (then dissipated in a turbulent cascade) produced by the engine, and are processes which re-equilibrate the energy balance of the climate system \cite{Lorenz67,Peix,Johnson,Luc09}.

A primary goal of climate science is to understand how the statistical properties - mean values, fluctuations, and higher order moments - of the climate system change as a result of modulations to some crucial external (e.g. solar irradiance) or internal (e.g. atmospheric composition) parameters of the system occurring on various time scales. A large class of problems - those involving climate sensitivity, climate variability, climate change, climate tipping points - fall into this category. In a system as complex and as extended as the climate, where lots of feedbacks are active on a variety of spatial and temporal scales, this is in general a very difficult task. The need for scientific advance in this direction is outstanding as one considers that even after several decades of intense scientific efforts, the accurate evaluation of the climate sensitivity \textit{par excellence}, i.e., the change of the globally averaged surface temperature for doubling of $CO_2$ concentration with respect to pre-industrial levels (280 ppm to 560 ppm circa), is a tantalizing endeavor, and large uncertainties are still present \cite{Ipcc2007}.

Such efforts have significant relevance also in the context of the ever-increasing attention paid by the scientific community to the quest for reliable metrics to be used for the validation of climate models  of various degrees of complexity and for the definition of strategies aimed at the radical improvement of their performance \cite{Held05,Luc08}. The pursuit of a \textit{quantum leap} in climate modelling - which definitely requires new scientific ideas rather than just faster supercomputers - is becoming more and more of a key issue in the climate community \cite{Shukla}.

A serious, fundamental difficulty in the analysis of the non equilibrium systems is that the fluctuation-dissipation relation \cite{Kubo2}, cannot be applied \cite{Ruelle1}. This greatly limits the ability of understanding the response of the systems to external perturbations by looking at its variability. In the specific case of climate, this can be rephrased by saying that climate change signals need not project on the natural modes of climate variability. The non-equivalence between free and forced climate fluctuations had been suggested by Lorenz \cite{Lorenz1}. The basic reason for this behavior is that, since the dynamics is forced and dissipative, with the asymptotic dynamics taking place in a strange attractor, natural fluctuations and forced motions cannot be equivalent. Whereas natural fluctuations of the system are restricted to the unstable manifold, because, by definition, asymptotically there is no dynamics along the stable manifold, external forcings will induce motions - of exponentially decaying amplitude - out of the attractor with probability one. The fluctuation-dissipation relation can be recovered only if we consider perturbations with the somewhat artificial property of being everywhere tangent to the unstable manifold or, in a more fundamental way, if we add a stochastic forcing, which has the crucial effect of smoothing the invariant measure \cite{Vul1,Vul2}. Potential links to these issues can be found in recent papers proposing new algorithms for three \cite{Trevi1} and four \cite{Trevi2} dimensional variational data assimilation, where it is shown that the quality of the procedure improves if the increment of the variables due to the assimilation is performed only along the unstable manifold.

Recently, Ruelle \cite{Ruelle1,Ruelle3} introduced a rigorous mathematical theory allowing for computing analytically,\textit{ ab initio}, the response of a large class of non-equilibrium systems to general external perturbations featuring arbitrary time modulation. The crucial result is that the changes in the expectation value of a physical observable can be expressed as a perturbative series in increasing powers of the intensity of the external perturbation, where each term of the series can be written as the expectation value of some well-defined observable over the unperturbed state. In a previous paper \cite{(Val08)} we showed that the Ruelle theory is, thanks to this property, formally analogous to usual Kubo response theory  \cite{Kubo1}, which applies for quasi-equilibrium system. The crucial difference lies on the mathematical properties of the invariant measure, which is absolutely continuous in the quasi-equilibrium case and singular in the non-equilibrium case.

Ruelle's analysis applies for non-equilibrium steady state systems \cite{Gallav2006} possessing a Sinai-Ruelle-Bowen (SRB) invariant measure, often referred to as Axiom A system \cite{EckRuelle,Ruelle2}. This class of systems, even if mathematically non-generic, includes on the other hand excellent models for general physical systems, as made clear by the chaotic hypothesis \cite{GallavCohen,Gallav}, which can be interpreted as an extension of the ergodic hypothesis to non-Hamiltonian systems \cite{Gallav2006}. See \cite{Penland} for an original geophysical perspective.

The Ruelle response theory, with the support of chaotic hypothesis, has interesting conceptual implications for climate studies. In fact, the possibility of defining a response function basically poses the problem of climate response to forcings and of climate change in a well-defined context, and, when considering the procedures aimed at improving climate models, justifies rigorously the procedures of tuning and adjusting the free parameters. Moreover, the response theory allows to compute the climate sensitivity, in the special case when static perturbations to the system parameters are considered.

Previously, a response formula was proposed by Cacuci for evaluating the linearized change of the solution of a time-independent generic system of nonlinear equations as a result of a change in the system's parameters \cite{Cacuci1,Cacuci2}. This can be interpreted as a special case of Ruelle's theory, where the unperturbed attractor is constituted by a fixed point and a static perturbation to the system evolution equation is considered. Cacuci proposed to study this problem using the adjoint operator to the original system, which provided an efficient way to determine the impact of small perturbations. Interestingly, early prominent applications of the so-called adjoint method and its extension to time-dependent problems, which allowed for evaluating all possible linear sensitivities of an evolving model in just one simulation, were been proposed for climate related problems. In particular, it was used to evaluate the sensitivities of a simple radiative-convective model possessing an attractor constituted by just one fixed point \cite{Hall82,Hall83}, and later, in an inherently heuristic way, for studying the response of a (chaotic) simplified general circulation model to doubling of the $CO_2$ concentration \cite{Hall86}. Whereas the adjoint method did not find much space in further climatic studies, mostly due to early discouragement for the computational burden of constructing the suitable operators for evaluating the sensitivities, it subsequently reached great success in data assimilation problems for geophysical fluid dynamics \cite{Ghil,Errico}, to the point that a tangent and adjoint model compiler able to automatically generate adjoint model code was has been introduced \cite{Giering1998}. More recently, a link between advanced adjoint techniques and the Ruelle theory has been proposed \cite{Lea}.

In the last decade on one side a great effort has been directed at extending the Ruelle response theory for more general classes of dynamical systems (see, e.g., \cite{Dolgo,Baladi}), and recent studies \cite{(Val08)} have shown that, thanks only to the causal nature of the response, it is possible to apply all the machinery of the theory of Kramers-Kronig (KK) relations \cite{Nussen,PeipVartAsakura,ValBass1,LSP05} for linear and nonlinear processes to study accurately and rigorously the susceptibilities describing in the frequency domain the response of a general observable to a general perturbation.

Moreover, the actual applicability of the theory has been successfully tested in a number of simple dynamical systems case for the linear \cite{Reick,CessSep} and nonlinear \cite{(Val09)} response. Such numerical investigations have clarified that even in systems which are not Axiom A, like the Lorenz 63 system \cite{Lorenz4}, it is possible to successfully use the response theory to construct linear \cite{Reick} and nonlinear susceptibilities \cite{(Val09)} which obey all of the constraints imposed by the KK theory up to a high degree of precision.

These investigations definitely motivate further studies aimed at understanding to what extent the response theory is an efficient tool for analyzing complex \textit{and} complicated systems. In this paper, we take up such a challenge and consider the Lorenz 96 (L96) system \cite{Lorenz3,LorenzEmm,Lorenz2}, which provides an excellent and celebrated prototypical model of a one dimensional atmosphere. The variables of the L96 model can be thought as generic meteorological quantities extending around a latitudinal circle and sampled at a regular interval. In spite of not being realistic in the usual sense, the L96 model presents the basic ingredients, such as dissipation, advection and the presence of an external forcing, of the actual atmosphere. For this reason, L96 has quickly become the standard model to be used for predictability studies \cite{Orrell,Haven,Hallerberg}, when testing data assimilation techniques \cite{Trevi1,Trevi2,Fertig}, and new parameterizations \cite{Wilks}.

Although we are unable to prove that the unperturbed L96 is an Axiom A system, in general and for the specific choice of parameters used in our numerical simulations in particular, we adopt the chaotic hypothesis and present the first thorough investigation of a spatially extended system by using the rigorous statistical mechanical methodologies presented in \cite{Ruelle1,Ruelle3,(Val08),(Val09)}. Moreover, since L96 is a spatially extended system, we  also explore the applicability of the response theory in all possible combinations of global/local observables and global/local perturbations. We compute rigorously the corresponding linear susceptibilities, verify the KK relations and the related sum rules, and find an empirical power law, which, as in \cite{LSV07}, supports the validity of the chaotic hypothesis, allowing to extend the results obtained for our specific choice of model's parameters to a rather general class of L96 systems. We also show how to go from the frequency back to the time domain, thus deriving from the susceptibility the Green function, which acts as time propagator of the considered perturbation for the considered observable. The Green function allows to predict, in an ensemble mean sense, the change in the observable at any time horizon as a result of a perturbation with the same spatial patter as that considered in the calculation of the susceptibility but featuring a general time modulation. 

Finally, building upon the results presented here, we propose a simple yet general methodology to study general Climate Change problems on virtually any time scale by resorting to only few, well selected simulations, and by taking full advantage of ensemble methods. The specific case of globally averaged surface temperature response to a general pattern of change of the $CO_2$ concentration is discussed.

Whereas the paper aims at proposing new methods for tackling classical problems of climate science, most of the results and of the methodologies proposed are of more general interest. In this paper we limit our attention to the linear response. We refer to \cite{(Val08),(Val09)} for a theoretical and numerical studies of higher-order effects of perturbations.

The paper is organized as follows. In Sec. \ref{theory} we briefly analyze the general theoretical background of the linear response theory and of the properties of the frequency dependent susceptibility and present some new useful results. In Sec. \ref{model} we present the main features of the L96 system, introduce the considered perturbations to the forcing, derive some basic properties of the response of various observables, and present the theoretical predictions. In Sec. \ref{resu} we present the results of our numerical investigations and describe how they can be generalized to the entire family of L96 models. In Sec. \ref{change} we provide a relevant example to illustrate how the results presented in this paper can be used to devise simple yet rigorous methods to study the climate response at all time scales on models of any degree of complexity. In Sec. \ref{conclu} we discuss the conclusions and present perspectives for future work.

\section{Theoretical Background: Ruelle Theory and Dispersion Relations}
\label{theory}
\subsection{Definition of the Linear Susceptibility}
We consider an Axiom A dynamical system described by the evolution equation $\dot{x}=F(x)$, so that the invariant probability measure
$\rho_0$ of the associated flow is of the SRB type \cite{Ruelle1}. Let $\left\langle \Phi\right\rangle_0$ be the expectation value of
the general observable $\Phi$ defined as $\int \rho_0(\textrm{d}x)\, \Phi(x)$. We perturb the flow of the system by adding a on the right hand side of the evolution equation a vector field $X(x)f(t)$, where $X(x)$ defines the pattern of the perturbation, and $f(t)$ is its time modulation. The resulting evolution equation results to be $\dot{x}=F(x)+X(x)f(t)$. Following Ruelle \cite{Ruelle1}, we express the expectation value of $\Phi(x)$ in the perturbed system using a perturbative expansions as:
\begin{equation}
\left\langle \Phi \right\rangle(t)= \left\langle \Phi \right\rangle_0 +\sum_{n=1}^{\infty} \left\langle \Phi\right\rangle^{(n)}(t).
\end{equation}
Each term of the perturbative series can be expressed as an n-convolution integral of the n$^{th}$ order causal Green function with n delayed time perturbation functions \cite{Ruelle1b,(Val08)}. Limiting our attention at the linear case we have:
\begin{equation}
\left\langle \Phi\right\rangle^{(1)}(t)=
 \int^{+\infty}_{-\infty}\textrm{d}\sigma_1 G_{\Phi}^{(1)}(\sigma_1)f(t-\sigma_1)
\label{risposta}
\end{equation}
The first order Green function can be expressed as follows:
\begin{equation}
\label{FirstGreen}
G^{(1)}_\Phi(\sigma_1)=\int \rho_0(\textrm{d}x)\Theta(\sigma_1)\Lambda\Pi_0(\sigma_1)\Phi(x),
\end{equation}
where $\Lambda(\bullet)=X(x)\nabla(\bullet)$ takes into account the effects of the perturbative vector field,  $\Theta$ is the usual Heaviside distribution and $\Pi_0$ the unperturbed time evolution operator so that $\Pi_0 F(x)=F(x(t))$ for any function $F$, with $x(t)$ following the unperturbed flow. Note that it is possible to express the Green function as the expectation value of a non-trivial but computable observable over the unperturbed SRB measure $\rho_0$. Therefore the knowledge of the unperturbed features of the flow is sufficient to define the effects of any external perturbation over any observable of our system. In the frequency domain we find that the first term of the perturbative series can be written as:
\begin{equation}
\label{Susciforce}
\left\langle \Phi\right\rangle^{(1)}(\omega)= \int^{+\infty}_{-\infty} \textrm{d}\omega_1  \chi^{(1)}_\Phi(\omega_1) f(\omega_1) \times \delta (\omega-\omega_1)= \chi^{(1)}_\Phi(\omega) f(\omega),
\end{equation}
where the Dirac delta implies that we are analyzing the impact of perturbations in the frequency-domain at the frequency $\omega$. The linear susceptibility is defined as:
\begin{equation}
\label{Gichi}
\chi_\Phi^{(1)}(\omega)= \int^{+\infty}_{-\infty} \textrm{d}t G^{(1)}_\Phi (t) \exp[\textrm{i}\omega t].
\end{equation}
It is important to underline with a thought experiment the computational relevance of the last equations and the importance of the susceptibility function.

Let's suppose we introduce a time dependent perturbation $f_\alpha(t)$ to a given pattern of forcing $X(x)$, simulate the system  and observe the time response of an arbitrary observable $\langle \Phi_\alpha\rangle^{(1)}(t) $. We now compute the Fourier transform of the observed signal and of the forcing modulation. Inverting Eq.\eqref{Susciforce}, we can find the linear susceptibility $\chi_\Phi^{(1)}(\omega)$. Let's now consider a different time-modulating function of the forcing $f_\beta(t)$ and its corresponding Fourier transform $f_\beta(\omega)$. Taking into account Eq. \eqref{Susciforce}, if we multiply $f_\beta(\omega)$ times the previously computed function $\chi_{\bar{\Phi}}^{(1)}(\omega)$ we directly obtain $\langle \Phi_\beta \rangle^{(1)}(\omega)$, the frequency-dependent response of the observable $\Phi$ to the forcing $X(x)$ modulated by the new function. By applying the inverse Fourier transform we obtain the time-dependent response $\langle \Phi_\beta\rangle^{(1)}(t) $ without needing any additional simulation.

Moreover, the knowledge of the susceptibility function allows us to reconstruct the $G_{\Phi}^{(1)}(t)$ by inverting Eq. \eqref{Gichi}. Otherwise, the Green function can be obtained directly from observing the response signal by performing a simulation where $f(t)=\delta(t)$: in this case (see Eq. \eqref{risposta}) we simply have $\left\langle \Phi\right\rangle^{(1)}(t)=G_{\Phi}^{(1)}(t)$.

\subsection{Kramers-Kronig relations and sum rules}
As we see from Eq. \eqref{FirstGreen}, for an arbitrary choice of the observable and of the perturbation the corresponding linear Green function is causal. Assuming, on heuristic physical basis, that $G_\Phi^{(1)}(t)\in L^2$, we can apply the Titchmarsh theorem \cite{Nussen,PeipVartAsakura,ValBass1,LSP05} and deduce that the linear susceptibility $\chi^{(1)}_\Phi(\omega)$ is a holomorphic function in the upper complex $\omega$-plane and the real and the imaginary part of $\chi(\omega)$ are connected to each other by Hilbert transform.

According to a general property of Fourier transform we know that the short term behavior of $G_\Phi^{(1)}(t)$ determines the asymptotic properties of $\chi_\Phi^{(1)}(\omega)$. We shall obtain a more quantitative result by exploiting that:
\begin{equation}
\int^{+\infty}_{-\infty}\textrm{d}t \Theta(t) t^k \exp[i \omega t] = (-i)^k\frac{\textrm{d}}{\textrm{d}\omega}\left(P\frac{\textrm{i}}{\omega}+\pi\delta(\omega)\right)\approx k! \frac{\textrm{i}^{(k+1)}}{\omega^{(k+1)}}
\end{equation}
where in the second equality we have neglected the fact that the solution is a distribution and considered $\omega\neq 0$. Therefore, if the Taylor expansion of the Green function in the limit $t\rightarrow 0^+$ is of the form:
\begin{equation}
\label{ShortGreen}
G^{(1)}_\Phi(t)\approx \bar{\alpha} \, \Theta(t) \, t^{\beta}  + o(t^\beta)
\end{equation}
the high frequency behavior of the linear susceptibility, i.e. the limit $\omega \rightarrow \infty$, is:
\begin{equation}
\label{asymomega}
\chi^{(1)}_\Phi(\omega)\approx \alpha \,\omega^{-\beta-1} + o(\omega^{-\beta-1})
\end{equation}
where $\alpha=\bar{\alpha}i^{(\beta+1)}\beta!$. The parameters $\beta$ (which is an integer number) and $\bar{\alpha}$ depend on the observable $\Phi$, on the specific features of the unperturbed system, and on the forcing under consideration. Taking into account \eqref{Gichi} and assuming that $\omega$ is real, we obtain that $\chi^{(1)}_\Phi(\omega)=[\chi^{(1)}_\Phi(-\omega)]^*$, so that Re$[\chi]$ is an even function while Im$[\chi]$ is odd function of $\omega$. Thus $\alpha=\alpha_R$ is real if $\beta$ is odd, whereas $\alpha=\textrm{i}\alpha_I$ is imaginary if $\beta$ is even.

Taking into account the Titchmarsch theorem, using that  $\chi^{(1)}_\Phi(\omega)=[\chi^{(1)}_\Phi(-\omega)]^*$, and considering the asymptotic behavior of the susceptibility, it is possible to show that the real and imaginary part of the linear susceptibility obey the following set of general KK dispersion relations\cite{LSP05}:
\begin{equation}
\label{ImKK}
-\frac{\pi}{2}\omega^{2p-1}\textrm{Im}[ \chi^{(1)}_\Phi(\omega) ] = P \int^{\infty}_{0} \textrm{d}\nu \frac{\nu^{2p}\textrm{Re}[ \chi^{(1)}_\Phi(\nu)] }{\nu^2-\omega^2}
\end{equation}
\begin{equation}
\label{ReKK}
\frac{\pi}{2}\omega^{2p}\textrm{Re}[ \chi^{(1)}_\Phi(\omega)] = P \int^{\infty}_{0} \textrm{d}\nu \frac{\nu^{2p+1} \textrm{Im}[ \chi^{(1)}_\Phi(\nu)] }{\nu^2-\omega^2}
\end{equation}
with $P$ indicating integration in principal part and $p=0,\ldots,(\beta-1)/2$ if $\beta$ is odd and $p=0,\ldots,\beta/2$ if $\beta$ is even. Note that the faster the asymptotic decrease of the susceptibility, the higher the number of independent constraints due to KK relations it has to unavoidably obeys. As thoroughly discussed in \cite{Kubo2}, in the case of quasi-equilibrium system, the fluctuation-dissipation theorem ensures that the imaginary part of the susceptibility describing the response of a given observable to a perturbation is proportional to a suitably defined power spectrum in the unperturbed system. Therefore, observing the unperturbed system and using Eq.\eqref{ReKK} it is possible to reconstruct the entire linear susceptibility, and so know everything about the response properties of the system. In the case of a non-equilibrium system, as discussed in the Introduction, this procedure is not possible.

It is possible to use the KK relations to define specific self-consistency properties of the real and imaginary part of the susceptibility. We first consider the following application: we set $p=0$ in Eq. \eqref{ReKK} and take the limit $\omega \rightarrow 0$. We obtain that for any observable:
\begin{equation}
\label{Staticsusc}
\textrm{Re}[\chi^{(1)}(0)]= \frac{2}{\pi} P \int \textrm{d}\nu \, \frac{\textrm{Im}[\chi^{(1)}( \nu)]}{\nu},
\end{equation}
which says that the static susceptibility (i.e., in a more common language, the linear sensitivity of the system) is related to the out-of-phase response of the system at all frequencies. In other terms, Eq. \eqref{Staticsusc} is an exact formula for the linear susceptibility of the system. Note that the static susceptibility is a real number because, thanks to the symmetry properties discussed above, Im$[\chi^{(1)}(0)]=0$. Moreover, we know that Re$[\chi^{(1)}(0)]$ is finite because the susceptibility function is analytic (and so in particular non singular). This is consistent with the fact that as $\omega\rightarrow 0$, the imaginary part of the susceptibility goes to zero at least as fast as a linear function, as only odd positive integer exponents can appear in its Taylor expansion around $\omega= 0$), so that the integrand in Eq. \eqref{Staticsusc} is not singular. Similarly, we obtain that for $\omega\sim 0$ the real part of the susceptibility is in general of the form $c_1+c_2\omega.^2+o(\omega^3)$, where $c_1$ and $c_2$ are two constants and $c_1$ is exactly given by Eq. \eqref{Staticsusc}.

By exploring the $\omega\rightarrow\infty$ limit in Eqs. \eqref{ImKK}-\eqref{ReKK} we obtain further integral constraints. By applying the superconvergence theorem \cite{Frye}, we obtain the following set of vanishing sum rules (see \cite{ValBass1}):
\begin{equation}
\label{vanishsum1}
\int^{\infty}_{0}  \textrm{d}\nu \nu^{2p+1} \textrm{Im}[ \chi^{(1)}_\Phi(\nu)]\,  = 0 \qquad   \begin{array} {c c} 0 \leq p \leq \beta/2-1, &\qquad \beta \qquad \mbox{even} \\ 0\leq p \leq (\beta-3)/2,& \qquad \beta \qquad \mbox{odd} \end{array}.
\end{equation}
\begin{equation}
\label{vanishsum2}
\int^{\infty}_{0} \nu^{2p} \textrm{Re}[ \chi^{(1)}_\Phi(\nu)] \, \textrm{d}\nu = 0 \qquad \begin{array} {c c} 0\leq p \leq \beta/2-1, \qquad & \beta \qquad \mbox{even} \\ 0\leq p \leq (\beta-1)/2, \qquad & \beta \qquad \mbox{odd} \end{array}.
\end{equation}
Note that if $\beta=0$ no vanishing sum rules can be written for the susceptibility, whereas if $\beta=1$ only Eq. \eqref{vanishsum2} provides a zero-sum constraint. For each set of KK relations, an additional, non-vanishing sum rule can be obtained. If $\beta$ is odd, the non-vanishing sum rule is:
\begin{equation}
\label{evensumrule}
\int^{\infty}_{0}  \nu^{\beta} \textrm{Im}[ \chi^{(1)}_\Phi(\nu)]\, \textrm{d}\nu  =-\frac{\pi}{2} \alpha_R,
\end{equation}
while if $\beta$ is even, we have:
\begin{equation}
\label{oddsumrule}
\int^{\infty}_{0} \nu^{\beta} \textrm{Re}[ \chi^{(1)}_\Phi(\nu)] \, \textrm{d}\nu =\frac{\pi}{2} \alpha_I,
\end{equation}
where the $\alpha$ constants are defined in Eq. \eqref{asymomega}. These sum rules provide additional general constraints that must be obeyed by any system and can be used to test the quality of the output of any model wishing to describe it. If we are not in the perfect model scenario (e.g., we use a simplified representation of some degrees of freedom) the sum rules can in principle be used to provide a fit for the parametrization.

We underline that it is possible to generalize the KK theory for specific classes of nonlinear susceptibilities for both quasi-equilibrium and non-equilibrium systems. Such results, which are particularly suited for studying the fundamental properties of harmonic generation processes, are thoroughly discussed in \cite{(Val08)} and will not be reported here.

\subsection{A practical formula for the linear susceptibility and consistency relations between susceptibilities of different observables}
As discussed above, the definition of the linear susceptibility does not depend on the function $f(t)$ modulating  the additional forcing, so that it is possible to draw general conclusions on its properties even by choosing a specific function $f(t)$.

Let's consider $f(t)= 2\epsilon\cos(\omega t)$. The impact of the perturbation on the evolution of a general observable $\Phi(x)$ is defined as:
\begin{equation}
\label{efpert}
\delta\Phi_\epsilon(t,t_0,x_0)=\Phi_\epsilon(t,t_0,x_0)-\Phi_0(t,t_0,x_0)
\end{equation}
where $x_0$ and $t_0$ are the initial condition and the initial time, and we associate the lower index $\epsilon$ to the strength of the forcing. The Ruelle's response theory ensures that $\delta\Phi_\epsilon(t,t_0,x_0)=O(\epsilon)$. Following \cite{Reick,(Val09)} the linear susceptibility results to be:
\begin{equation}
\label{chivera}
\chi_\Phi^{(1)}(\omega)\equiv \lim_{\epsilon \rightarrow 0}\lim_{T \rightarrow \infty}\chi_\phi^{(1)}(\omega,x_0,\epsilon,T)
\end{equation}
where:
\begin{equation}
\chi_\Phi^{(1)}(\omega,x_0,\epsilon,T)=\frac{1}{T} \, \frac{1}{\epsilon} \int^{T}_{0}\textrm{d}t\delta\Phi_{\epsilon}(t,x_0) \exp(\textrm{i}\omega t)
\label{chifinita}
\end{equation}
is the total susceptibility, affected by the finite time and finite size response of the system. This quantity depends on the initial condition and in principle contains information about the response of the system at all order of nonlinearity.

Since  $\textrm{d}(\delta\Phi_{\epsilon}(t,x_0))/\textrm{d}t = \delta  (\textrm{d}\Phi_{\epsilon}(t,x_0))/\textrm{d}t$, thanks to the linearity of the time derivative, by considering Eqs. \eqref{chivera}-\eqref{chifinita} and performing an integration by parts, we obtain that \cite{Reick}:
\begin{equation}
\chi_{\dot{\Phi}}^{(1)}(\omega)=-\textrm{i}\omega \chi_{\Phi}^{(1)}(\omega).
\label{chiderivata}
\end{equation}
Let's now find a different expression for $\chi_{\dot{\Phi}}^{(1)}(\omega)$. The time derivative of $\Phi$ in the unperturbed system is
\begin{equation}
\label{Phidotu}
\dot{\Phi}(x)=\Gamma(x),
\end{equation}
where $\Gamma=F\cdot \nabla \Phi$. Similarly, the time derivative for $\Phi$ in the case of the perturbed motion described by  $\dot{x}=F(x)+X(x)f(t)=F(x)+2\epsilon \cos (\omega t)X(x)$ is:
\begin{equation}
\label{Phidotp}
\dot{\Phi}(x)=\Gamma(x)+2\epsilon \cos (\omega t)\Xi(x),
\end{equation}
where $\Xi= X \cdot \nabla \Phi$. From Eqs. \eqref{Phidotu}-\eqref{Phidotp} we obtain that
\begin{equation}
\delta \dot{\Phi}_{\epsilon}(t,x_0)=\delta \Gamma_\epsilon(t,x_0)+2\epsilon \cos (\omega t)\Xi(t,x_0),
\label{pertphidot}
\end{equation}
where all terms are of $O(\epsilon)$. Furthermore, we integrate each term in Eq. \eqref{pertphidot} as in Eq. \eqref{chifinita}, take the limits $\epsilon\rightarrow 0$ and $T\rightarrow \infty$, and obtain:
\begin{align}
\nonumber
\lim_{\epsilon \rightarrow 0}\lim_{T \rightarrow \infty}\frac{1}{T}& \, \frac{1}{\epsilon} \int^{T}_{0}\textrm{d}t\delta\dot{\Phi}_{\epsilon}(t,x_0)  \exp(\textrm{i}\omega t) =\lim_{\epsilon \rightarrow 0}\lim_{T \rightarrow \infty}\frac{1}{T} \, \frac{1}{\epsilon} \int^{T}_{0}\textrm{d}t\delta{\Gamma}_{\epsilon}(t,x_0)\exp(\textrm{i}\omega t)+\\
&+ \lim_{\epsilon \rightarrow 0}\lim_{T \rightarrow \infty}\frac{1}{T} \, \frac{1}{\epsilon} \int^{T}_{0}\textrm{d}t \epsilon \Xi_{\epsilon}(t,x_0) \left[\exp(\textrm{i}\omega t)+\exp(-\textrm{i}\omega t)\right]\exp(\textrm{i}\omega t).
\end{align}
Using the definition in Eq. \eqref{chivera} and the identity given in Eq. \eqref{chiderivata}, we derive:
\begin{align}
\label{chicons1}
-\textrm{i}\omega \chi_{\Phi}^{(1)}(\omega) & = \chi_{\Gamma}^{(1)}(\omega) )
+ \lim_{\epsilon \rightarrow 0}\lim_{T \rightarrow \infty}\frac{1}{T} \, \int^{T}_{0}\textrm{d}t \Xi_{\epsilon}(t,x_0)+\nonumber \\
& + \lim_{\epsilon \rightarrow 0}\lim_{T \rightarrow \infty}\frac{1}{T}\frac{1}{\epsilon} \, \int^{T}_{0}\textrm{d}t \epsilon \Xi_{\epsilon}(t,x_0)\exp(2\textrm{i}\omega t).
\end{align}
The first limit in Eq. \eqref{chicons1} gives, by definition, $\langle \Xi \rangle_0$, whereas the second limit vanishes as the expression under integral is $O(\epsilon^2)$, since it is related to second order harmonic generation nonlinear process \cite{(Val09)}. Concluding, we obtain the following general consistency relation for the linear susceptibility:
\begin{equation}
\label{chiconsfinale}
\chi^{(1)}_\Phi(\omega) = \frac{\textrm{i}}{\omega}\chi^{(1)}_\Gamma(\omega)+\frac{\textrm{i}}{\omega}\langle \Xi \rangle_0.
\end{equation}
Such an identity related the susceptibility of an observable $\Phi$ to the susceptibility of the projection of its gradient along the unperturbed flow $\Gamma$ and to the average value in the unperturbed state of the projection of its gradient along the perturbation flow. Note that the two terms on the right hand side are radically different. Whereas the first term is related to the projection of the dynamics along the unstable manifold, the second term depends on the structure of the forcing $X(x)$, which may be entirely unrelated to that of the unstable manifold. This is the fundamental reason why the fluctuation-dissipation theorem does not apply in the non-equilibrium case.

Moreover, since, as shown in Eq. \eqref{ShortGreen}-\eqref{asymomega}, the susceptibility of a generic observable decreases to zero at least as fast as $\omega^{-1}$, for large values of $\omega$ we have that $\chi^{(1)}_\Phi(\omega) \approx {i}/\omega \langle \Xi \rangle_0$ unless $\omega\langle \Xi \rangle_0=0$. If $\omega\langle \Xi \rangle_0\neq0$, we also have that the leading order of the short-time expansion of the Green function is of the form:
\begin{equation}
\label{Greenshort2}
G^{(1)}_\Phi(t) = \Theta(t) \langle \Xi \rangle_0 +o(t^0),
\end{equation}
in agreement with what can be found by direct inspection of Eq. \eqref{FirstGreen}.
\section{Application of the Response Theory to the Lorenz 96 Model}
\label{model}
\subsection{Statistical properties of the unperturbed Lorenz 96 Model}
\label{L96testo}
The Lorenz 96 model \cite{Lorenz3,LorenzEmm,Lorenz2} describes the evolution of a generic atmospheric variable defined in N equally spaced grid points along a latitudinal circle  and provides a simple, unrealistic but conceptually satisfying representation of some basic atmospheric processes, even if such one-dimensional model it cannot be derived \textit{ab-initio} from any dynamic equation via subsequent approximations.  The evolution equations can be written in a scaled form as follows:
\begin{equation}
\label{L96}
\frac{\textrm{d}\,x_i}{\textrm{d}\,t}=x_{i-1}(x_{i+1}-x_{i-2})-x_i+F
\end{equation}
where $i=1,2,.....,N$, and the index $i$ is cyclic so that $x_{i+N}=x_{i-N}=x_{i}$. The quadratic term in the equations simulates advection, the linear one represents thermal or mechanical damping and the constant one is an external forcing. The evolution equations are invariant under $ i\rightarrow i+1 $, so that the dynamics is the same for all variable.  The time scale of the system is given by the damping time, which corresponds to five days. The L96 system shows different features, as different choices of $F$ and $N$ may strongly alter the topology of the attractor, alternating periodic, quasi-periodic and chaotic behavior in a non trivial way. However with a suitable choice of the parameters $N$ and $F$, the system is markedly chaotic. In particular, as $F$ controls the energy input into the system, we expect that for relatively high values of this parameter the system should simulate a turbulent behavior and live on a strange attractor.  As an example, setting $N=40$ and $F=8$ the system possesses 13 positive Lyapunov exponents, the largest corresponding to a doubling time of 2.1 days, while the fractal dimension of the attractor \cite{Ruelle2} is about 27.1 \cite{LorenzEmm}.

When computing the time derivative of the total energy of the system, defined as $E=1/2\sum_ix^2_i$, the advection  terms cancel. The evolution equation for $E$ results to be:
\begin{equation}
\label{energydot}
\dot{E} = -2E + F\sum_i x_i.
\end{equation}
As the dynamics takes place inside a compact set, $\Psi(x_i,\ldots,x_N)$ is bounded for any choice of the function $\Psi$. Therefore the ensemble mean with respect the $\rho_0$ (or time average) of the temporal derivative $\dot{\Psi}$  vanishes. Therefore, defining $M=\sum_i x_i$ as the total momentum of the system, we obtain the following identity:
\begin{equation}
\label{energy}
2\left\langle E \right\rangle_0 = \sum_i\left\langle  x_i^2 \right\rangle_0=F\sum_i\left\langle  x_i \right\rangle_0= F\left\langle  M\right\rangle_0.
\end{equation}
Similarly, we can deduce an additional consistency relation by investigating the expression of the time derivative of M:
\begin{equation}
\label{Mequazio}
\left\langle M\right\rangle_0= NF + \left\langle C_2\right\rangle_0-\left\langle C_3\right\rangle_0
\end{equation}
where $\left\langle C_2\right\rangle_0=\sum_i\left\langle x_i x_{i-2}\right\rangle$ and
$\left\langle C_3\right\rangle_0=\sum_i\left\langle x_ix_{i-3}\right\rangle_0$. Higher order consistence relations can be obtain in a similar fashion.

The equivalence of all the variables implies that over the unperturbed flow each observable $O$ of the whole system satisfies $\sum_i\left\langle  O(x_i) \right\rangle_0= N \left\langle O(x_j)  \right\rangle_0$ $\forall j$. Therefore, we define the average energy per grid point $e(N,F)$ and the average momentum per grid point $m(N,F)$ as:
\begin{equation}
\label{asymen}
e\,=\, \frac{\left\langle x_i^2\right\rangle_0}{2}\,= \,\frac{ \left\langle  E\right\rangle_0}{N},
\end{equation}
\begin{equation}
\label{asymmed}
m\,= \,\left\langle x_i\right\rangle_0\,=\, \frac{ \left\langle M\right\rangle_0 }{N},
\end{equation}
where the choice of $i$ is arbitrary and the $N$ and $F$-dependence is dropped for shortness. Defining $c_i=\left\langle C_i \right\rangle_0/N$, $c_0=2e$, we can rewrite Eqs. \eqref{energy}-\eqref{Mequazio} as follows:
\begin{equation}
\label{intensiven}
2\,e= F\, m,
\end{equation}
\begin{equation}
\label{intensivmed}
m= F+c_2-c_3.
\end{equation}
Expressing either the average energy  $e$ or the average momentum $m$ per grid point as a function of the two free parameters $N$ and $F$ would allow to get a closure equation for the statistical properties of the unperturbed Lorenz 96 system. We have computed $e(N,F)$ and $m(N,F)$ by performing long integrations for values of $F$ ranging from 6 to 50 with step 1 and for values of $N$ ranging from 10 to 200 with step 10. In all of these cases, chaotic motions are observed. We have consistently found that, within 0.5\%, $e(N,F)=e(F)$ and $m(N,F)= m(F)$, so that they can be considered intensive quantities. Therefore, we can interpret Eqs. \eqref{intensiven}-\eqref{intensivmed} as equations providing a definition of the thermodynamics of this simple one-dimensional model of atmosphere.

The $F$-dependence of $e$ and $m$ can be closely approximated in terms of simple power laws. We obtain, within a precision of about 1\% in the considered domain, that
\begin{equation}
m(F)=\lambda F^ {\gamma},
 \end{equation}
and, consistently with Eq.(\ref{intensiven}),
\begin{equation}
e(F)=\frac{\lambda}{2} F^{1+\gamma},
\end{equation}
with $\lambda \approx 1.15$ and $\gamma \approx 0.35$. Such a smooth dependence of the intensive energy and momentum with respect to the forcing parameter $F$ is indeed in agreement with the hypothesis that the invariant measure is deformed in a very regular fashion not only locally, but over a large range of the parameter's space.

Note that, at the fixed point of the system corresponding to a purely zonally symmetric dynamics ($x_i=F$, $\forall i)$ we have $m=F$ and $e=F^2/2$. These formulas give much higher values for both $m$ and $e$ than what found with our empirical power laws for the attractor in the chaotic regime. In fact, at such an equilibrium, which is unstable in the parametric range explored here, the energy dissipation is much weaker than in the co-existing chaotic attractor, which corresponds to the case where breaking nonlinear waves and turbulent motions are present. Interestingly, the presence of well-defined scaling laws with respect to the forcing parameters for the energy and momentum of the system with different characteristic exponents in the chaotic regime and in the co-existing unstable equilibrium is in agreement with previous finding recently obtained in a simple baroclinic quasi-geostrophic model \cite{LSV07}.

\subsection{Asymptotic properties of the linear susceptibility}
\label{asym}
We perturb the L96 model by adding a small perturbation modulated by $f(t)= 2\epsilon\cos(\omega t)$. The resulting evolution equation is:
\begin{equation}
\label{L96pert}
\frac{\textrm{d}\,x_i}{\textrm{d}\,t}=x_{i-1}(x_{i+1}-x_{i+2})-x_i+F+2\epsilon\cos(\omega t)X_i
\end{equation}
where $X_i=X_i(x_1, \ldots, x_N)$ is a generic function of variables $x_i$. We adopt the chaotic hypothesis \cite{Gallav,GallavCohen} and we follow the theory proposed by Ruelle \cite{Ruelle1} and discussed in Sect. \ref{theory} in order to study the linear response of suitably defined observables to the perturbation. We first propose to study the high-frequency, response by analyzing in detail the asymptotic properties of the resulting susceptibilities. As discussed in section \ref{theory}, this constitutes a crucial step for constructing the set of applicable KK relations and for computing the value of the sum rules.

We consider two different forcing patters $X_i$. In the first case, we apply the perturbation over all the grid points and we choose $X_i=1 \, \forall \, i$. Given an observable $\Phi$, we refer to the linear Green function and linear susceptibility resulting from this choice of $X_i$ as $G_{\Phi,N}^{(1)}$ and $\chi_{\Phi,N}^{(1)}$, respectively, where the lower index $N$ indicates that the perturbation acts over all the variables $x_i$. We refer to this pattern of forcing as \textit{global perturbation}. In the second case, we apply the perturbation only on the variable $x_j$ of the L96 model, and we choose $X_1=0 \forall i\neq j$, $X_j=1$, . Since all the points are equivalent in the unperturbed case, the choice of $j$ is arbitrary. In this case, when referring to the linear Green function and the linear susceptibility, the lower index $1$ substitutes the $N$, indicating that the perturbation is localized to one point. We refer to this pattern of forcing as \textit{local perturbation}.

\subsubsection{Global perturbation}
\label{glope}
We consider perturbations with spatial pattern given by $X_i=1 \, \forall \, i$  and analyze the response of the observable $E$. Following Eq. \eqref{FirstGreen}, the linear Green function $G_{E,a}^{(1)}(t)$ can be explicitly written as:
\begin{align}
\label{ge}
G_{E,N}^{(1)}(t) &= \int \rho_0(\textrm{d}x) \Theta(t)\Lambda\Pi_0(t)E(\vec{x}) \nonumber \\
&=\int \rho_0(\textrm{d}x) \Theta(t) \,\vec{1}\cdot \vec{\nabla} E(\vec{x}(t))\nonumber \\
& = \int \rho_0(\textrm{d}x) \Theta(t)\sum_i \partial_i (E(\vec{x}(t)))
\end{align}
where $\vec{x}(t)$ satisfies the unperturbed evolution Equation \eqref{L96}. Taking into account Eq. \eqref{ShortGreen} and Dq. \eqref{asymomega}, in order to obtain the asymptotic behavior of the susceptibility, we need to study the short time behavior of the Green function. Therefore, we express $E(\vec{x}(t))$ as a Taylor series about $t=0$ considering the unperturbed flow, compute the integral of each coefficient of the $t$-expansion over $\rho_0$, and seek the lowest order non-vanishing term \cite{(Val09)}. The first two terms of the Taylor expansion of $E$ in Eq. \eqref{ge} give:
\begin{align}
G_{E,N}^{(1)}(t) & = \int \rho_0(\textrm{d}x) \Theta(t) \sum_i \partial_i \, \Big(E|_{t=0}+t\dot{E}|_{t=0}+o(t)\Big)\nonumber \\
& = \int \rho_0(\textrm{d}x)\Theta(t) \Big[ \sum_i x_i - \Big(\sum_i 2x_i - NF\Big)t + o(t)\Big].
\end{align}
Using Eqs. \eqref{ShortGreen}-\eqref{asymomega}, the leading terms of the asymptotic behavior of linear susceptibility can be written as:
\begin{align}
\label{chientutti}
\chi^{(1)}_{E,N}(\omega)&=i\Big(\sum_i\left\langle x_i \right\rangle_0\Big)/\omega + \Big( \sum_i\left\langle 2x_i\right\rangle_0- NF \Big)/\omega^2 +o(\omega^{-2}) \nonumber \\
&= iN\frac{m}{\omega} -N\frac{F-2m}{\omega^2} +o(\omega^{-2})
\end{align}
Since the symmetry with respect the index $i$ is valid also in the perturbed case, given our choice of the forcing pattern, the linear susceptibility of the total energy is given the sum of $N$ identical contributions, each corresponding to the susceptibility of the observable $\varepsilon=1/2x_i^2$ for each of the N variables $x_i$ of the system. Therefore, it is possible to define an intensive linear susceptibility  $\chi^{(1)}_{\varepsilon,N}=1/N\chi^{(1)}_{E,N}$, where $\chi^{(1)}_{\varepsilon,N}$ describes the response of the local energy to the external perturbation. In particular in the limit $\omega \rightarrow\infty$ we have:
\begin{equation}
\label{chiensingle}
\chi^{(1)}_{\varepsilon,N}(\omega)= i\, \frac{m}{\omega} - \frac{F-2m}{\omega^2} +o(\omega^{-2}).
\end{equation}
Equations \eqref{chientutti} and \eqref{chiensingle} imply that the imaginary part dominates the asymptotic behavior of the susceptibility, so that at high frequency the response is shifted by about $\pi/2$ with respect the forcing. Observing that the leading term of asymptotic $\chi$ is of order $\omega^{-1}$ just one sum rules apply for either susceptibilities. Limiting our attention to the intensive quantity $e$, by applying Eq. \eqref{oddsumrule} we obtain:
\begin{equation}
\label{sumrulensingle}
\int^{\infty}_{0} \textrm{Re}[\chi_{\varepsilon,N}^{(1)}(\omega)]\, \textrm{d}\omega = \frac{\pi}{2} m.
\end{equation}
Along similar lines, if we select as observable the total momentum $M$, we derive that the asymptotic behavior of its linear susceptibility is:
\begin{equation}
\label{chimomtutti}
\chi^{(1)}_{M,N}(\omega)= N \,\chi^{(1)}_{\mu,N}(\omega)=\textrm{i}\frac{N}{\omega}+\frac{N}{\omega^2}+o(\omega^{-2}),
\end{equation}
where we have defined the intensive susceptibility $\chi^{(1)}_{\mu,N}(\omega)$, where $\mu$ is the intensive momentum of the system.  As in Eq. \eqref{chiensingle}, the asymptotic behavior is determined by the imaginary part of $\chi$, and the real part of the susceptibility provides the following sum rule:
\begin{equation}
\label{sumrulmedsingle}
\int^{\infty}_{0}  \textrm{Re}[\chi_{\mu,N}^{(1)}(\omega)]\, \textrm{d}\omega= \frac{\pi}{2}.
\end{equation}
We now wish to go back to the general consistency equation for linear susceptibilities given in Eq. \eqref{chiconsenemom}. Considering that in the perturbed system the time derivative of the total energy of the system can be written as:
\begin{equation}
\label{energydotn}
\dot{E} = -2E + FM+2\epsilon\cos(\omega t)M
\end{equation}
the general result given in Eq. \eqref{chiconsfinale} can be written as follows:
\begin{equation}
\label{chiconsenemom}
\chi^{(1)}_{E,N}(\omega) = \frac{F}{2-\textrm{i}\omega}\chi^{(1)}_{M,N}(\omega)+\frac{1}{2-\textrm{i}\omega}\langle M \rangle_0,
\end{equation}
since in this case $\Gamma=-2E+FM$ and $\Xi=M$. It is easy to check that the asymptotic behavior for the susceptibilities given in Eqs. \eqref{chientutti}-\eqref{chimomtutti} is in agreement with Eq. \eqref{chiconsenemom}, which is valid at all frequencies.
\subsubsection{Local perturbation}
\label{lope}
We now perturb the system in a single grid point. The symmetry of the unperturbed system implies the equivalence of every point of the latitude circle. Indicating with $x_j$ the grid point where forcing is exerted, the pattern of the perturbation vector field is $X_1=0 \forall i\neq j$, $X_j=1$. We consider the same modulating monochromatic function $f(t)=2\epsilon \cos(\omega t)$ as in the previous case. We analyze the asymptotic behavior of the linear susceptibility of the total energy of the system $E$, of the total momentum of the system $M$, which are global variables, and of the local variables constituted by the energy $E_j=1/2x_j^2$ and momentum $M_j=x_j$ of the perturbed grid point and of its immediate neighbors.

Since we are looking at the linear response and the global perturbation is given by $N$ spatially shifted copies of the local perturbation, for any observable $\Phi$ of the form $\Phi(x_1,\ldots,x_n)=\sum_{i=1}^N\phi(x_i)$ we must have $\chi^{(1)}_{\Phi,1}=\chi^{(1)}_{\Phi/N,N}=\chi^{(1)}_{\phi,N}$.

In the case of the observable $E$, it is straightforward to verify the previous identity at least in the asymptotic regimes. In fact, the short time behavior of the Green function describing the response of the $E$ to the local perturbation results to be:
\begin{align}
G_{E,1}^{(1)}(t) &= \int \rho_0(\textrm{d}x) \Theta(t) \partial_j \, \Big(E|_{t=0}+t\dot{E}|_{t=0} + o(t)\Big)\nonumber \\
  & = \int \rho_0(\textrm{d}x)\Theta(t)\partial_j \Big[ \frac{1}{2}\sum x_i^2 - \Big(\sum 2x_i^2 - x_iF\Big)t\Big]\nonumber \\
  & = \Theta(t) \Big(\left\langle 2x_j\right\rangle_0-\left\langle 2x_j -F\right\rangle_0t + o(t)\Big),
\end{align}
so that the asymptotic behavior of the corresponding linear susceptibility susceptibility is:
\begin{equation}
\label{enuno}
\chi^{(1)}_{E,1}=i\,\frac{m}{\omega} -  \frac{F-2m}{\omega^2} +o(\omega^{-2}),
\end{equation}
which agrees with what found for the intensive energy response when the global perturbation is applied (see Eq. \eqref{chiensingle}). The sum rule for the real part of the susceptibility is exactly the same as in what given in \eqref{sumrulensingle}:
\begin{equation}
\label{sumrulentotonepert}
\int^{\infty}_{0} \textrm{Re}[\chi_{E,1}^{(1)}(\omega)]\, \textrm{d}\omega = \frac{\pi}{2}.
\end{equation}
Analogously, we obtain that the asymptotic behavior of $\chi_{M,\epsilon,1}^{(1)}$ can be written as:
\begin{equation}
\chi_{M,1}^{(1)}(\omega) = \textrm{i} \frac{1}{\omega} +\frac{1}{\omega^2},
\end{equation}
with the corresponding sum rule:
\begin{equation}
\label{Summedone}
\int_{0}^{\infty} \textrm{Re}[\chi_{M,1}^{(1)}(\omega)]\, \textrm{d}\omega = \frac{\pi}{2},
\end{equation}
in perfect agreement with Eqs. \eqref{chimomtutti} and \eqref{sumrulmedsingle}, respectively.

It is rather interesting to look into local energy observables. Considering the energy $E_j$ of the perturbed grid point $x_j$ we have that its short term Green function can be written as:
\begin{align}
G_{E_j,1}^{(1)}(t) & = \int \rho_0(\textrm{d}x) \Theta(t) \partial_j \,\Big[ \frac{1}{2} x_1^2+ \Big(x_j(x_{j-1}x_{j+1}-x_{j-1}x_{j-2}-x_j+F)\Big)t +o(t)\Big] \nonumber \\
 & = \Theta(t) \Big[\left\langle x_j\right\rangle_0-\left\langle x_{j-1}x_{j+1}-x_{j-1}x_{j-2}-x_j+F\right\rangle_0t +o(t)\Big].
\end{align}
Since
\begin{equation}
\label{timedev}
0=\left\langle\dot{x_j} \right\rangle_0= \left\langle x_{j-1}x_{j+1}-x_{j-1}x_{j-2}-x_j+F \right\rangle_0
\end{equation}
because the $\rho_0$-average of the temporal derivative of any observable vanishes, thanks to the compactness of the attractor, we obtain the following asymptotic behavior for the linear susceptibility
\begin{equation}
\label{chienuno}
\chi^{(1)}_{E_j,1}=i\, \frac{m}{\omega} +\frac{m}{\omega^2}+o(\omega^{-2}).
\end{equation}
Since, by linearity, $\chi^{(1)}_{E,1}=\sum_k \chi^{(1)}_{E_k,1}$, comparing this result with what obtained in Eq. \eqref{enuno}, we note that the susceptibility of the energy at the position of the forcing $E_j$ provides the leading asymptotic term to the susceptibility of the total energy $E$. Consequently, in the high-frequency range $\chi^{(1)}_{E_j,1}\approx\chi^{(1)}_{E,1}$, and the two susceptibilities obey the same non-vanishing sum rule, so that:
\begin{equation}
\label{sumrulenoneonepert}
\int^{\infty}_{0} \textrm{Re}[\chi_{E_j,1}^{(1)}(\omega)]\, \textrm{d}\omega =  \frac{\pi}{2}m
\end{equation}
Nevertheless, by comparing Eqs. \eqref{enuno}-\eqref{chienuno}, we discover that contributions to the second leading order ($\propto\omega^{-2}$) in the high frequency range of the susceptibility of the total energy do not come just from the response of the energy at the perturbed grid point. perturbed grid point but some other point give a contribution of order $\omega^{-2}$. Therefore, the asymptotic behavior of the real part of $\chi^{(1)}_{E,1}$ is not captured by $\chi^{(1)}_{E_j,1}$. The locality of the interaction suggests to look at the energy of the closest neighbors of $x_j$. Because of the asymmetry of the nonlinear terms in the L96 evolution equations, we consider the observable $\psi=1/2(E_{j+1}+E_{j+2}+E_{j-1})$. It is possible to prove that:
\begin{equation}
\chi^{(1)}_{\psi,1}(\omega)=- \frac{\left\langle x_{j-1}(x_{j+1}-x_{j-2}) \right\rangle_0}{\omega^2}=-\frac{(F-m)}{\omega^2} +o(\omega^{-2}).
\end{equation}
It is easy to observe that the sum of $\chi^{(1)}_{\psi,1}$ and $\chi^{(1)}_{E_j,1}$ provides the correct leading order to the asymptotic behavior of both the real and imaginary parts of $\chi^{(1)}_{E,1}$. We shall provide an argument why this strongly supports the close resemblance of the two functions $\chi^{(1)}_{E,1}$ and $\chi^{(1)}_{E_{loc},1}$, where $E_{loc}=E_j+\psi=1/2x_j^2+1/2x_{j+1}^2+1/2x_{j+2}^2+1/2x_{j-1}^2$ is the energy of the cluster of points centered in $x_j$.

The analysis of the asymptotic behavior of the susceptibilities related to the local momentum of the system provides additional insights. It is possible to prove that for large frequency the linear susceptibility of the momentum of the perturbed grid point is:
\begin{equation}
\chi_{x_j,1}^{(1)}(\omega) = \textrm{i} \frac{1}{\omega} +\frac{1}{\omega^2}+o(\omega^{-2}),
\end{equation}
which suggests that the response of the local momentum captures the correct asymptotic behavior of both the real and the imaginary part of the total momentum $M$. Concluding, we obtain that the following sum rule can be stablished:
\begin{equation}
\label{Summedone1}
\int_{0}^{\infty} \textrm{Re}[\chi_{x_j,1}^{(1)}(\omega)]\, \textrm{d}\omega = \frac{\pi}{2}.
\end{equation}
Therefore, such a constraint is exactly the same whether we analyze the the response of the momentum of a single variable when the perturbation acts over all the grid points, or of the total momentum in the case of a local perturbation, or, in this latter case, of the momentum of the grid point where the local perturbation is applied.

As we have seen in this section, the coefficients of the leading asymptotic terms and the sum rules are given by simple linear functions of $m$ (or equivalently, thanks to Eq. \eqref{intensiven}, by $e$) and by $F$. As we have proposed an efficient parameterization of $m$ and $e$ as functions of $F$ alone in subsection \ref{L96testo}, our results can be easily applied and numerically verified for a very large class of L96 models.

\section{Results}
\label{resu}
\subsection{Simulations and Data Processing}
The accurate calculation of the linear susceptibility of the general observable $\Phi$ is not as easy task, since the definition given in Eq. (\ref{chivera}) requires the evaluation of two limits, whereas we can actually compute only the quantity given in Eq. (\ref{chifinita}). Averaging the response over a long time $T$ allows for improving the signal-to-noise ratio. Noise is present because, due to the chaotic nature of the flow, we have a continuous spectral background. Instead, considering small values for the perturbation strength $\epsilon$ degrades the signal-to-noise ration, but, on the other hand, it is crucial to select a small $\epsilon$ in order to keep the perturbations as close as possible to the linear regime. As discussed in \cite{(Val09)}, we can improve the signal-to-noise ratio without needing to perform very long integrations and  to consider large values for $\epsilon$ by performing an ergodic averaging of the quantity averaging the quantity $\chi_\Phi^{(1)}(\omega,x_i,\epsilon,T)$. Therefore, we choose the best estimator of the true susceptibility $\chi_\phi^{(1)}$ as:
\begin{equation}
\chi_\Phi^{(1)}(\omega) \cong \lim_{K \rightarrow \infty}\frac{1}{K}\sum_{i=1}^K \chi_\Phi^{(1)}(\omega,x_i,\epsilon,T),
\label{practicalchi}
\end{equation}
where the $x_i$ are randomly selected initial conditions chosen on the attractor of the unperturbed system.

The numerical integrations of the Lorenz 96 system have been performed using the standard configuration where $N$, the number of degrees of freedom, is set to 40, and $F$, the intensity of the unperturbed forcing, is set to 8 \cite{Lorenz2,Lorenz3}. Equations \eqref{L96} and \eqref{L96pert} are solved using the standard fourth order Runge-Kutta numerical scheme.

For a given observable $\Phi$, the susceptibility at angular frequency $\omega$ is computed by applying Eq. (\ref{practicalchi}) to $K$ outputs of Eq. \eqref{L96pert}, each starting with a different initial condition, where the perturbation has the same angular frequency $\omega$. The angular frequency $\omega$  ranges from $\omega_l=0.2\pi$ to $\omega_h=20\pi$ with steps of $0.01\pi$. Each simulation performed with a perturbation of angular
frequency $\omega$ runs from $t=0$ up to $t=T=400\pi/\omega$, which corresponds to 200 full periods of the forcing. The length of the simulations depends on the corresponding period of the forcing because we are interested in obtaining a frequency-independent quality for the signal. We have observed that the linear response approximation is obeyed to a good degree of approximation for up to $\epsilon\approx 1$, which implies that the third order nonlinear effects are relatively small. See \cite{(Val08),(Val09)} for further clarifications on this latter point.

When considering the susceptibilities describing the response to the global perturbation, we present results obtained using $\epsilon=0.25$ and averaging over $K=100$ random initial conditions. When assessing the linear response to the local perturbation, a reasonably clear signal is obtained using $\epsilon=1$ and averaging over $K=300$ initial conditions.

Note that, since we are interested in the linear response, it is could have been possible to compute the susceptibility using a generic modulating function $f(t)$ (see Eq. (\ref{Susciforce})) rather than having to resort to multiple monochromatic perturbations. Nevertheless, for reasons of clarity, and for emphasizing that chaotic dynamical systems can be analyzed using tools typical of spectroscopy, we have used a more cumbersome but probably more convincing approach.

We underline that the numerical results have been obtained using a commercial laptop rather than resorting to HPC. This comes from the motivation of showing that the methodology presented is robust enough that relatively low-end means allow us to see the physical and mathematical properties of our interest. We emphasize that, using HPC, it is rather easy to greatly increase the quality of the signal by increasing $K$ and/or $T$ by a one or two orders of magnitude.
\subsection{Global Perturbation}
\begin{figure} 
   \begin{center}
   \includegraphics[width=0.5\textwidth,angle=270]{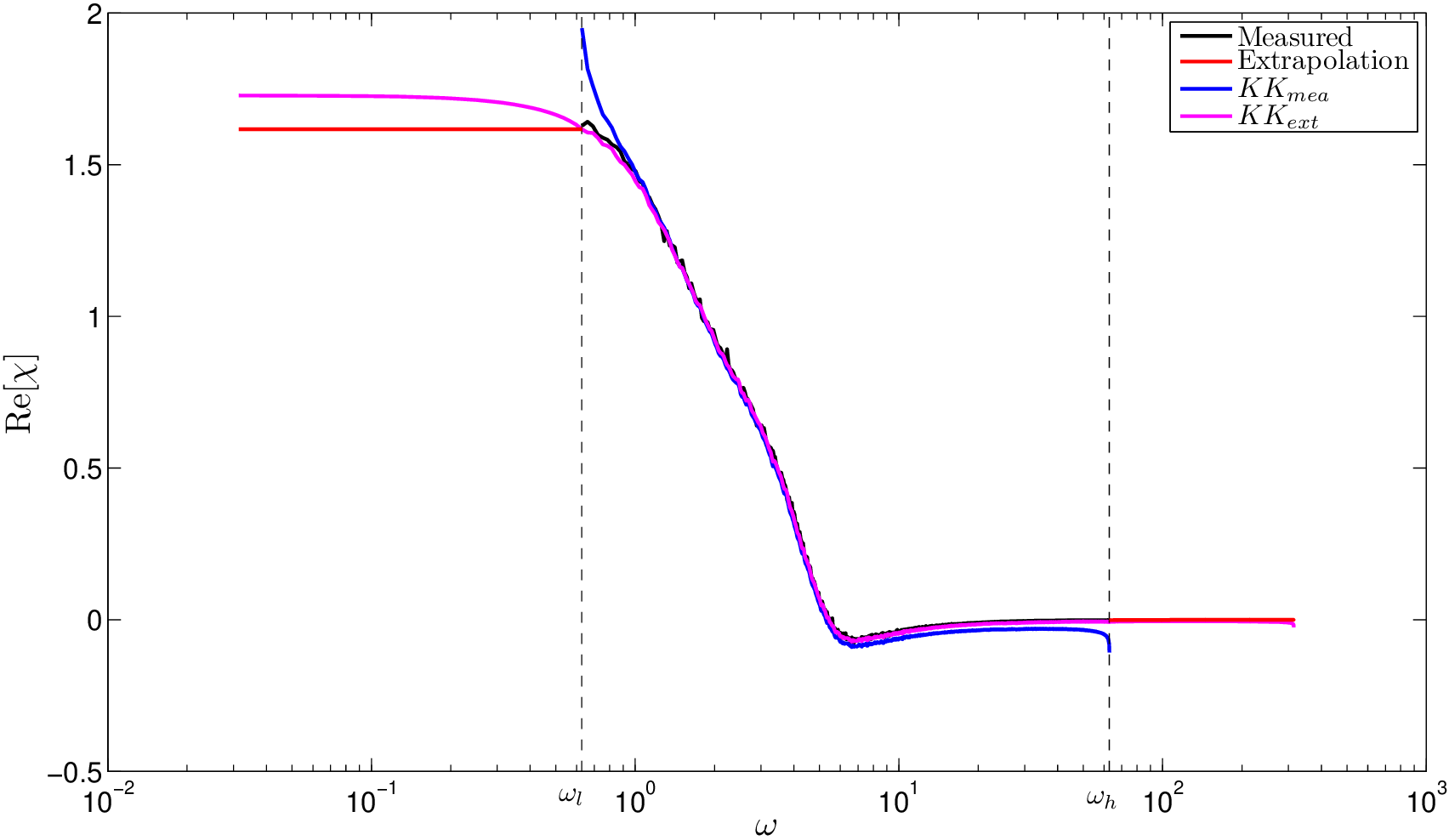}
   \includegraphics[width=0.5\textwidth,angle=270]{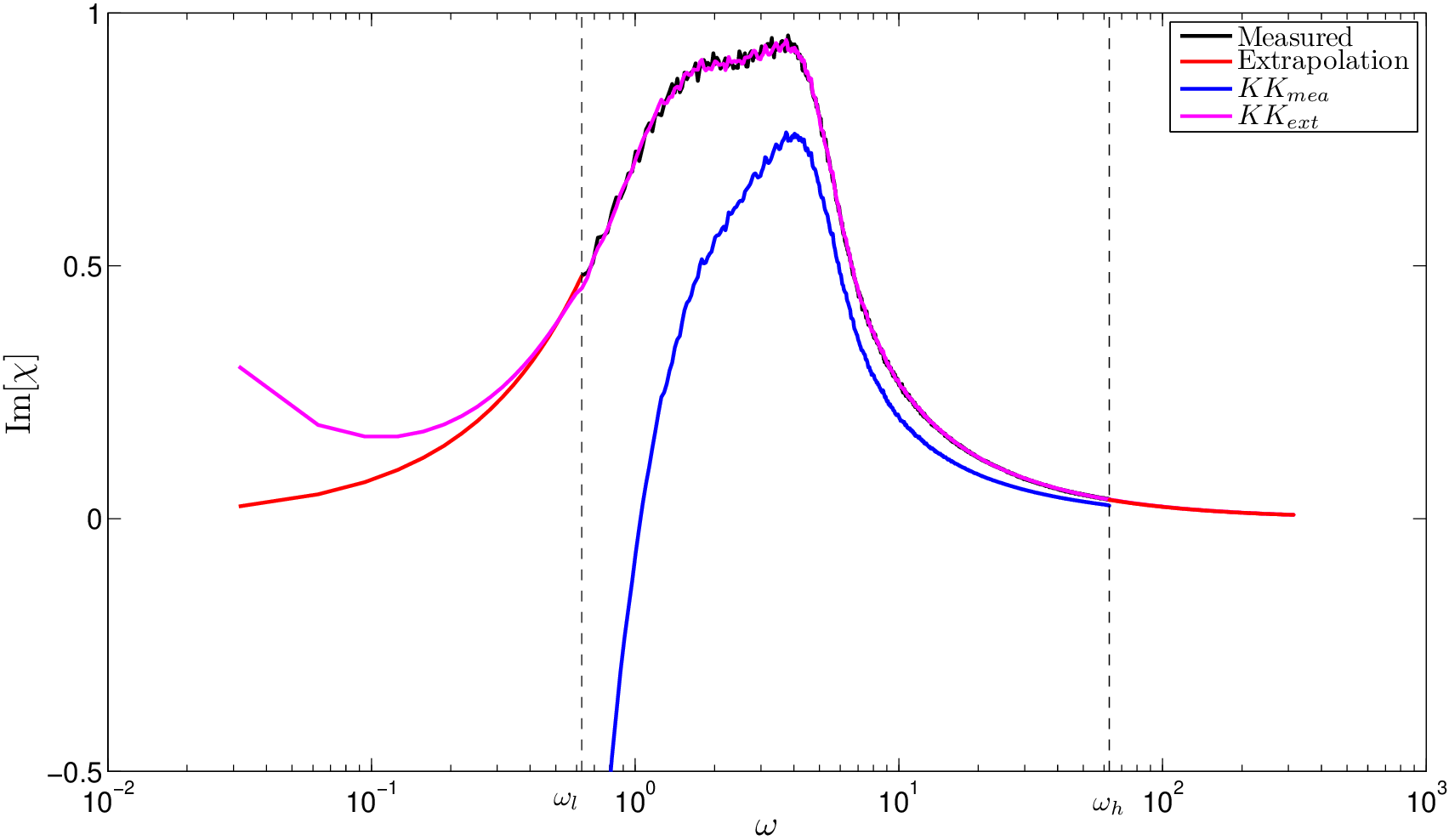}
      \end{center}
    \caption{Linear susceptibility of intensive energy of the system $E/N$ with respect to the global perturbation with $X_i=1$ $\forall i$. The real and the imaginary parts are depicted in Panels a) and b), respectively. The measured and extrapolated values are shown in red and black lines, respectively. The result of the Kramers-Kronig inversion done with the measured and with with the extrapolated data are shown in blue and magenta lines, respectively.}
  \label{chienergy}
\end{figure}

We first consider $\chi^{(1)}_{\varepsilon,N}=1/N \chi^{(1)}_{E,N}$, where $\varepsilon=E/N$, and follow up from the discussion in subsection \ref{glope}. The measured real and imaginary parts of the susceptibility are depicted with the black lines in Fig. (\ref{chienergy})a,b. The imaginary part has a broad spectral feature (with two distinct peaks) spanning from $\omega\approx 2$ to $\omega\approx 4$, which corresponds to about twice the time scale (=1) of the system and to four times (see Eq. \eqref{energydot}) the relaxation time of the energy. This hints at the fact that it is not obvious to constrain the spectral features of the response an observable just by performing a scale analysis of its evolution equation. For higher values of $\omega$, the imaginary part decreases in a very regular way, so that in the upper range a very good agreement with the asymptotic behavior $\sim m/\omega$ presented in Eq. (\ref{asymomega}) is obtained. For low frequencies, the imaginary part appears to decrease towards zero, as expected from symmetry reasons. Instead, the real part presents a dispersive structure in correspondence with the broad maximum of the imaginary part, and changes sign for $\omega\approx 6$, so that it is negative for high values of the frequency range. The asymptotic decrease to zero in this range is also in excellent agreement with the estimate $\sim - (F-2m)/\omega^2$ given in Eq. (\ref{asymomega}), whereas for low frequencies the real susceptibility tends to a very high value, this suggesting that the strongest response is obtained for static perturbations.

\begin{figure} 
   \begin{center}
   \includegraphics[width=0.5\textwidth,angle=270]{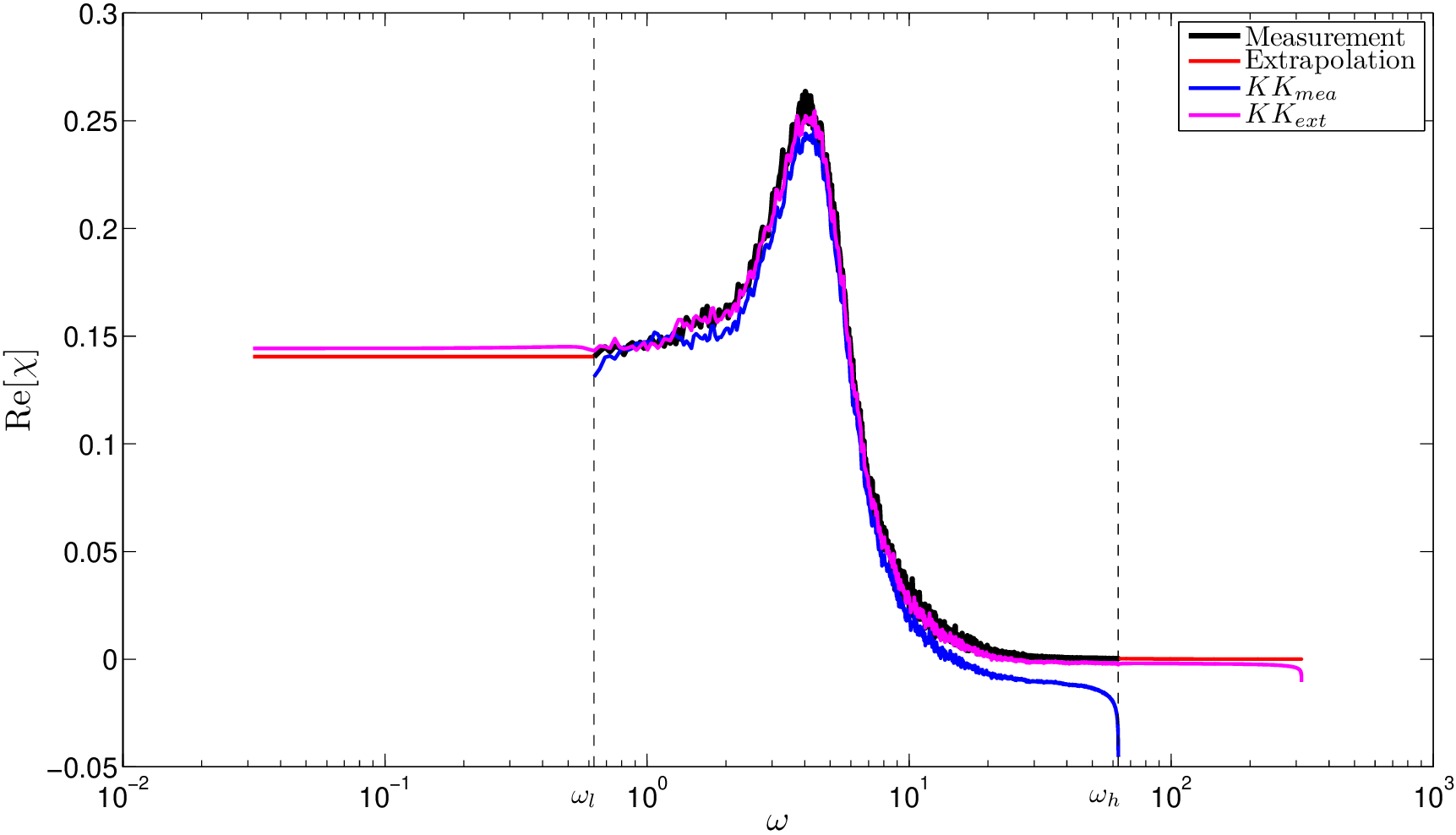}
   \includegraphics[width=0.5\textwidth,angle=270]{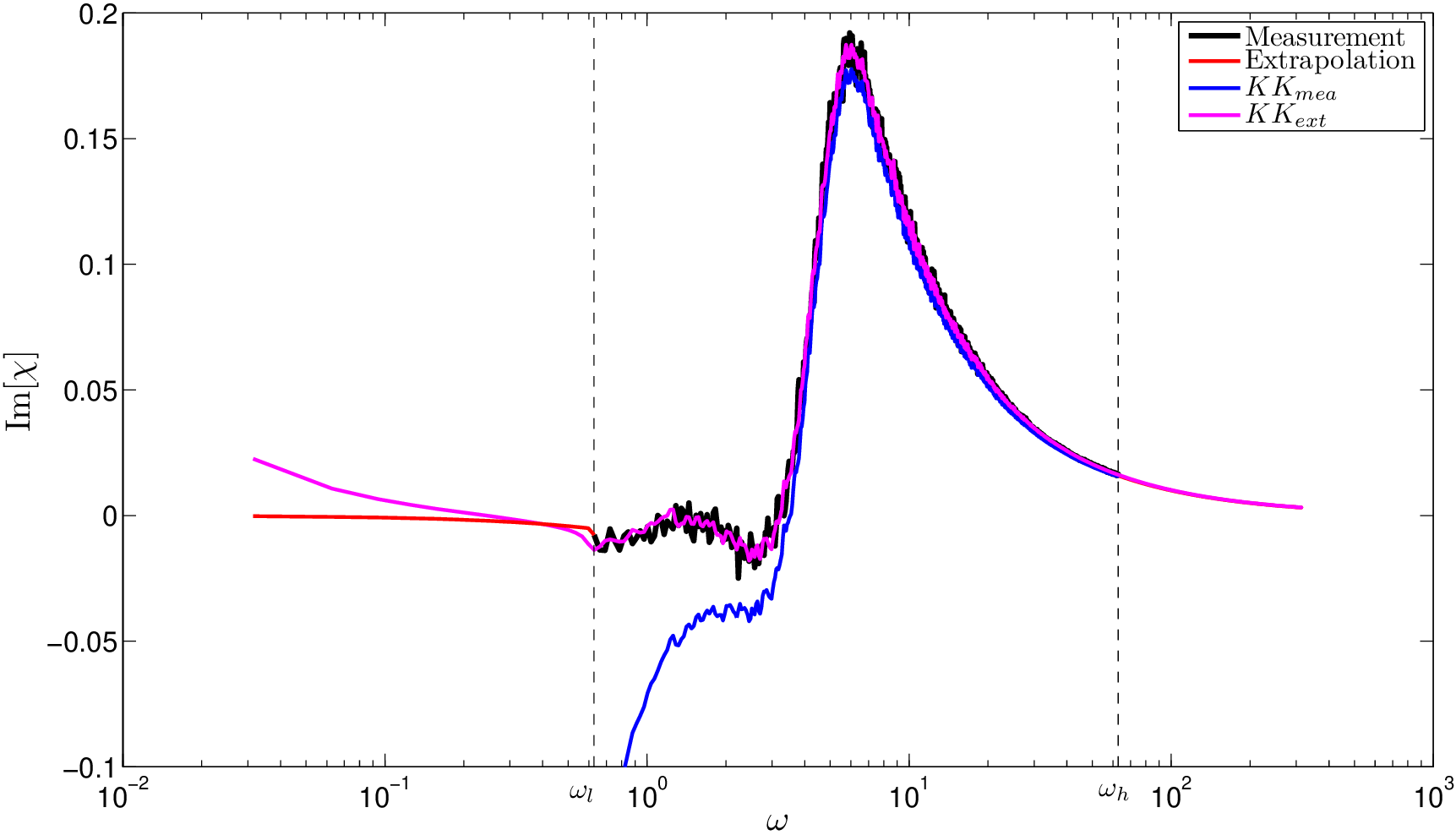}
      \end{center}
    \caption{Linear susceptibility of intensive momentum of the system $M/N$ with respect to the global perturbation with $X_i=1$ $\forall i$. The real and the imaginary parts are depicted in Panels a) and b), respectively. The measured and extrapolated values are shown in red and black lines, respectively. The result of the Kramers-Kronig inversion done with the measured and with with the extrapolated data are shown in blue and magenta lines, respectively.}
  \label{chimom}
\end{figure}

The measured real and imaginary parts of $\chi^{(1)}_{\mu,N}=1/N \chi^{(1)}_{M,N}$, where $\mu=M/N$, are depicted in black in Fig. (\ref{chimom})a,b. Interestingly, the spectral feature of the imaginary part is shifted to higher frequencies than in the case of the energy susceptibility, so that a well-distinct peak centered on value of $\omega\approx 6$, which approximately corresponds to the natural time scale of the system. For low frequencies, the susceptibility has almost exclusively a real component. As opposed to the previous case, the largest value for the in-phase response is not obtained for ultralow frequencies, but rather for $\omega\approx 4$. The asymptotic behavior of both the real and imaginary parts is in perfect agreement with the theoretical result given in Eq. (\ref{asymmed}), so that they are found to decrease asymptotically for high frequencies as $1/\omega^2$ and $1/\omega$, respectively.

We apply the truncated KK relations to the measured data to test the quality of the data inversion process. The estimates of the imaginary part (starting from the measured data of the real part) and of the real part (starting from the measured data of the imaginary part) obtained by applying Eqs. (\ref{ImKK})-(\ref{ReKK}) are shown for $\chi^{(1)}_{\varepsilon,N}$ in blue in Fig. (\ref{chienergy})a,b and for $\chi^{(1)}_{\mu,N}$ in Fig. (\ref{chimom})a,b. We observe that whereas agreement is very good for the real part for both susceptibilities, only a qualitative match is obtained for the imaginary part, with large discrepancies for $\omega\lesssim 2$. In this latter case, moreover, the well-known problem of KK divergence  at the boundaries of integration \cite{ValBass1,LSP05} is very serious for $\omega=\omega_l$.

It is crucial to test whether the discrepancies are due to the finiteness of the spectral range or are, instead, due to basic problems in the applicability of the Ruelle response theory, related to the fact that the invariant probability measure of the unperturbed system actually features large deviations from an SRB measure.

We proceed testing the first case. In order to widen the spectral range over which the susceptibility is defined, we will exploit the asymptotic properties obtained in section \ref{asym} as well as the low frequency behavior of the susceptibility discussed in section \ref{theory}. We redefine the the imaginary part of the susceptibility of $\chi^{(1)}_{\varepsilon,N}$ as follows
\begin{equation}
\label{imext}
\textrm{Im}[\chi^{(1)}_{\varepsilon,N}(\omega)]=\left\{{\begin{array}{l l}
 \frac{\omega}{\omega_l} \textrm{Im}[\chi^{(1)}_{\varepsilon,N}(\omega_l)],& \mbox{ } 0\leq\omega\leq \omega_l,\\
  \textrm{Im}[\chi^{(1)}_{\varepsilon,N}(\omega)], & \mbox{ }\omega_l\leq\omega\leq \omega_h, \\
  \frac{m}{\omega}, & \mbox{ } \omega\geq \omega_h,
  \end{array}}\right.
\end{equation}
where the measured data are sandwiched between the low and high frequency limit. Whereas we have a rigorous result for the high frequency limit, the low frequency limit is computed by making the reasonable assumption that the leading order of the $\omega\rightarrow 0$ limit is linear (see discussion after Eq. (\ref{Staticsusc})). Similarly, the real part of the susceptibility $\chi^{(1)}_{\varepsilon,N}$ can be redefined as follows:
\begin{equation}
\label{reext}
\textrm{Re}[\chi^{(1)}_{\varepsilon,N}(\omega)]=\left\{{\begin{array}{l l}
 \textrm{Re}[\chi^{(1)}_{\varepsilon,N}(\omega_l)],& \mbox{ } 0\leq\omega\leq \omega_l,\\
  \textrm{Re}[\chi^{(1)}_{\varepsilon,N}(\omega)], & \mbox{ }\omega_l\leq\omega\leq \omega_h, \\
  -\frac{F-2m}{\omega^2}, & \mbox{ } \omega\geq \omega_h,
  \end{array}}\right.
\end{equation}
where we have used the fact that at low frequencies the real part of the susceptibility is constant in $\omega$ up to a quadratic term. A corresponding procedure is used to extend the spectral range of the $\chi^{(1)}_{\mu,N}(\omega)$, where the suitable asymptotic behaviors described in subsection \ref{glope} are adopted. The red lines in Figs. (\ref{chienergy})a,b-(\ref{chimom}a,b) present the results of such extrapolations, and the magenta lines show the outcome of the data inversion of these functions performed via KK relations. We observe that the agreement is outstanding, with almost perfect overlap inside the region where measurement is performed and remarkable agreement also in the low and high frequency range. This is a very convincing evidence that the Ruelle response theory can be successfully applied for this system. Since the KK relations provide, first and foremost, consistency tests, the agreement the original and the KK-transformed susceptibility automatically confirms that the extrapolation procedure we have adopted is correct. A still better agreement would be found had we taken into account value of $\omega$ larger than what considered in the extrapolation used here (up to 100 $\pi$).

Furthermore, let's consider the results presented in subsection \ref{L96testo}. The slopes of the functions $e(F)$ and $m(F)$ are given by
\begin{equation}
\frac{\textrm{d}\,m(F)}{\textrm{d}\,F}=\lambda \gamma F^ {\gamma-1},
 \end{equation}
\begin{equation}
\frac{\textrm{d}\,e(F)}{\textrm{d}\,F}=\lambda\frac{(1+\gamma)}{2} F^{\gamma}=m(F)\frac{(1+\gamma)}{2}.
\end{equation}
They correspond, by definition, to the static susceptibility of the observables $e$ and $m$, respectively, for the global perturbation with $X_i=1$ considered here. When evaluating the derivatives of $e(F)$ and $m(F)$ for $F=8$ we obtain $(\textrm{d}\,e(F)/\textrm{d}\,F)_{F=8}\approx 1.6$ and $(\textrm{d}\,m(F)/\textrm{d}\,F)_{F=8}\approx 0.11$. These values are in good agreement with what found by extrapolating the corresponding real part of the susceptibilities for $\omega\rightarrow 0$ via KK relations and shown in Figs. (\ref{chienergy}a) and (\ref{chimom}a).

Apart from the verification of the validity of KK relations, we  want to provide further support for the quality of the linear susceptibilities considered.

\begin{figure} 
   \begin{center}
   \includegraphics[width=0.5\textwidth,angle=270]{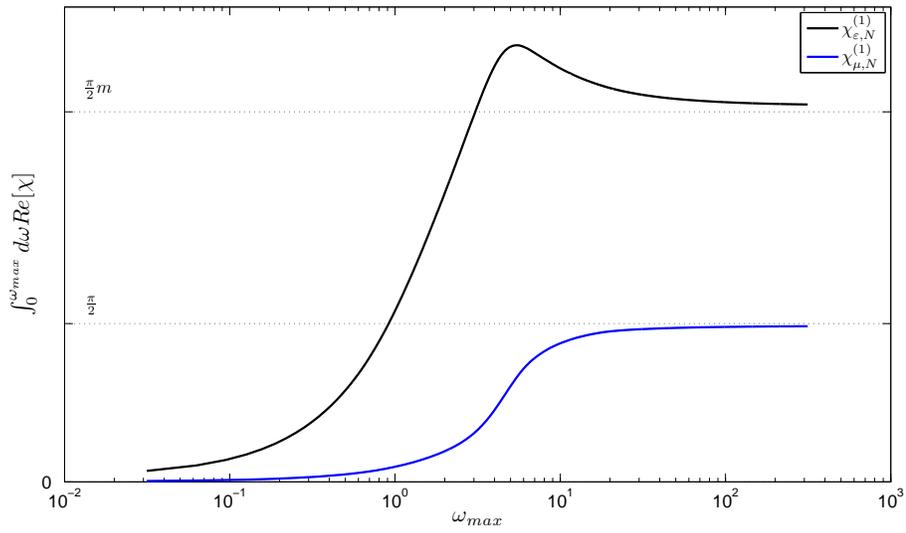}
      \end{center}
    \caption{Sum Rules for the real part of the linear susceptibility of intensive energy $E/N$ (black line) and momentum $M/N$ (blue line) of the system  with respect to the global perturbation with $X_i=1$ $\forall i$. The theoretical values are indicated in the figure.}
  \label{sumrulesenmom}
\end{figure}

First, we test the sum rules given in Eq. \eqref{sumrulensingle} and \eqref{sumrulmedsingle} for the real part of the extrapolated susceptibilities $\chi^{(1)}_{\varepsilon,N}(\omega)$ and $\chi^{(1)}_{\mu,N}(\omega)$, respectively. Our findings are presented in Fig. (\ref{sumrulesenmom}, where it is shown that an excellent agreement (within 1\%) is found between the theoretical values and the numerical results. Since Re$[\chi^{(1)}_{e,\epsilon,a}](\omega)$ is negative in the high-frequency range, the convergence of the integral to the theoretical value of the sum rule is from above, whereas the opposite occurs for Re$[\chi^{(1)}_{m,\epsilon,a}(\omega)]$. Extending the integral for even larger values of $\omega$ would bring the numerical results to an almost perfect agreement with the theory.

\begin{figure} 
   \begin{center}
   \includegraphics[width=0.5\textwidth,angle=270]{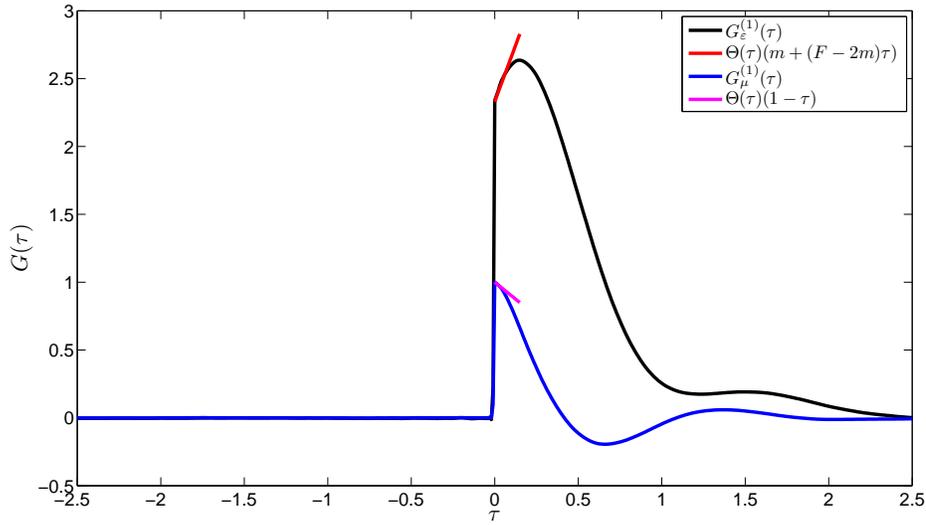}
      \end{center}
    \caption{Green functions describing the time-dependent response of the observable of intensive energy $E/N$ (black line) and momentum $M/N$ (blue line)  with respect to the global perturbation with $X_i=1$ $\forall i$. For both observables, the short time behavior of the Green function estimated from the asymptotic behavior of the corresponding susceptibility is shown in figure.}
  \label{green}
\end{figure}

Following the definition given in Eq. (\ref{FirstGreen}), the Green function $G^{(1)}_{\Phi}(\tau)$ computed for an observable $\Phi$ and a given pattern of perturbation flow $X_i(x)$ (in this case $X_i=1 \forall i$) can be used to compute the time-dependent linearized impact of all perturbations with the same spatial pattern $X_i(x)$ but with arbitrary time modulation. Whereas the direct estimate of the Green function from the time dependent dynamics can be obtained by performing an ensemble of simulations where the time modulation of the perturbation is given by a $\delta(t)$ pattern (see discussion in section \ref{theory}, we take the indirect route by considering Eq. (\ref{Gichi}). By applying the inverse Fourier Transform, we derive the Green functions corresponding to $\chi^{(1)}_{\varepsilon,a}(\omega)$ and $\chi^{(1)}_{\mu,a}(\omega)$. The results are presented in Fig. (\ref{green}): for both observables the Green functions are clearly causal, and their short-time behavior agrees remarkably well with what be deduced by looking at the asymptotic properties of the corresponding susceptibilities (compare Eqs. (\eqref{chiensingle} and \eqref{chimomtutti}).

\subsection{Local Perturbation}
The data obtained for the numerical simulations of the response to the local perturbation are, given the much weaker overall strength of the forcing, much noisier that those presented in the previous section. Nevertheless, we shall see that all the theoretical predictions are verified to a surprisingly good degree of approximation.

\begin{figure} 
   \begin{center}
   \includegraphics[width=0.5\textwidth,angle=270]{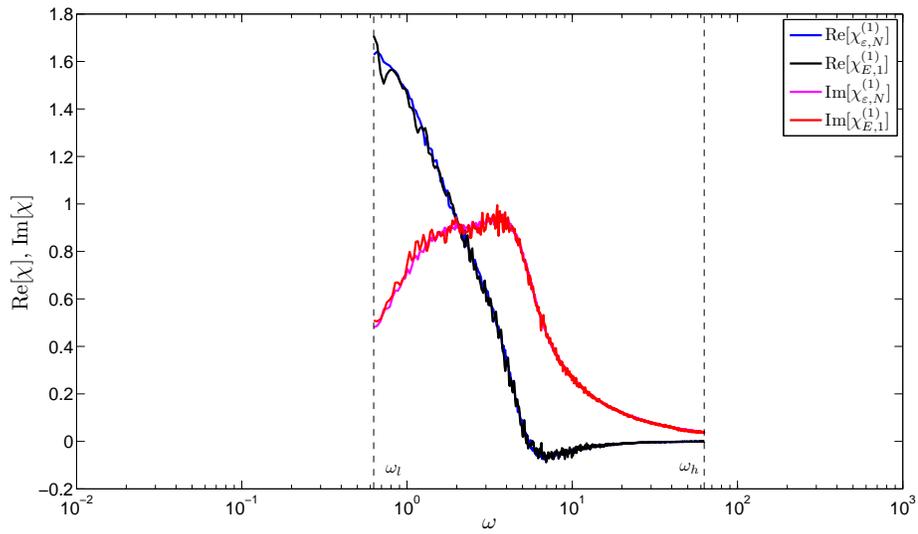}
      \end{center}
    \caption{Comparison between the linear susceptibility of the intensive energy for the global perturbation and of the total energy for the local perturbation. The signals are the same except for the different level of noise. }
  \label{tuttiuno}
\end{figure}

The global observables $E$ and $M$ are of the form $\Phi(x_1,\ldots,x_n)=\sum_{i=1}^N\Phi(x_i)$, where $\phi(x)=x^2/2$ and $\Phi(x)=x$, respectively. We have consistently verified that the identity $\chi^{(1)}_{\Phi,1}=\chi^{(1)}_{\Phi/N,N}=\chi^{(1)}_{\phi,N}$ discussed in the previous section applies in the whole spectral range explored by our simulations, compatibly with the (slightly) different signal-to-noise ratios in the two sets of simulations. See, e.g., Fig. \ref{tuttiuno} for the comparison between the two susceptibilities $\chi^{(1)}_{E,1}$ and $\chi^{(1)}_{\varepsilon,N}$.

\begin{figure} 
   \begin{center}
   \includegraphics[width=0.5\textwidth,angle=270]{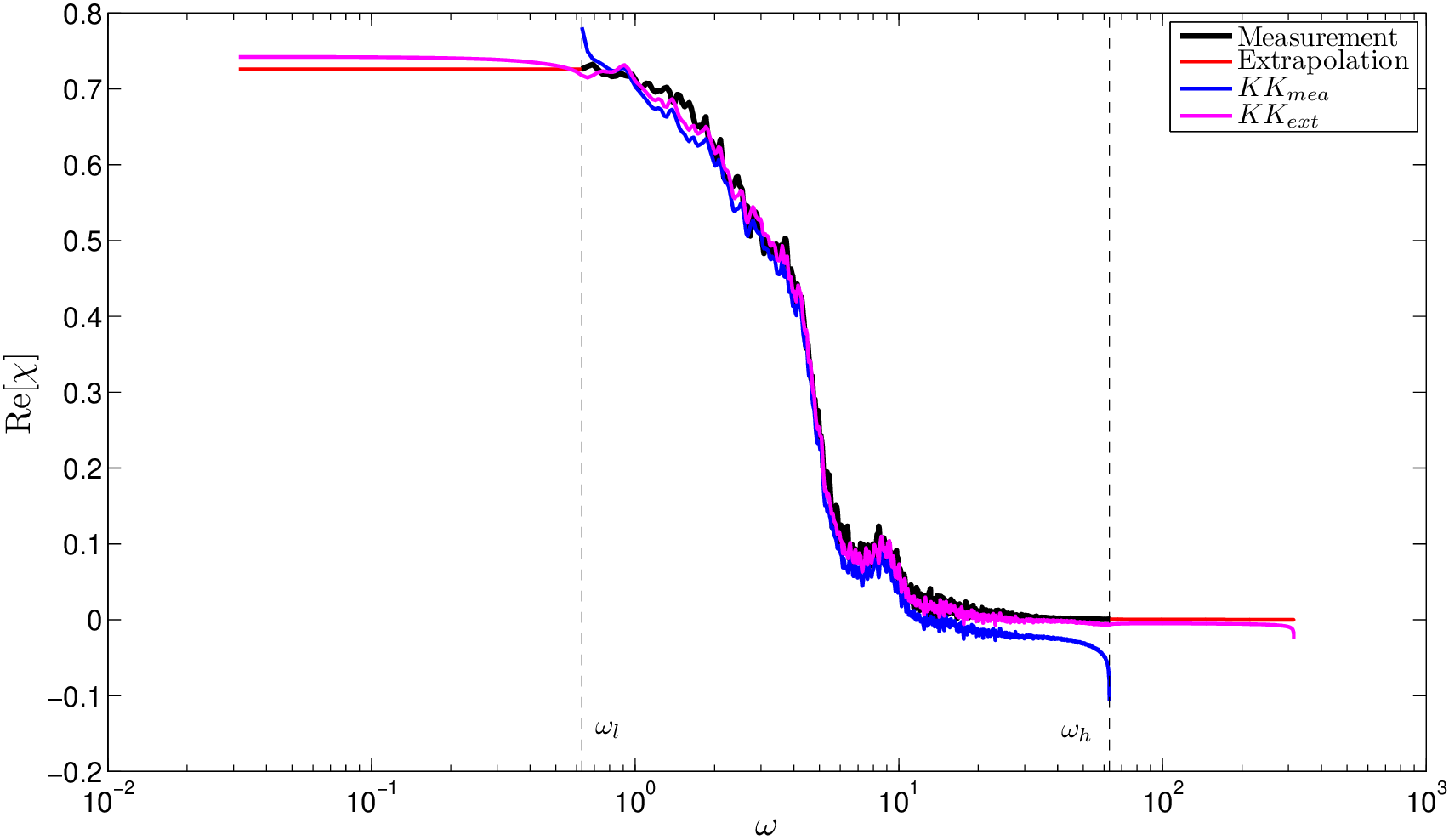}
   \includegraphics[width=0.5\textwidth,angle=270]{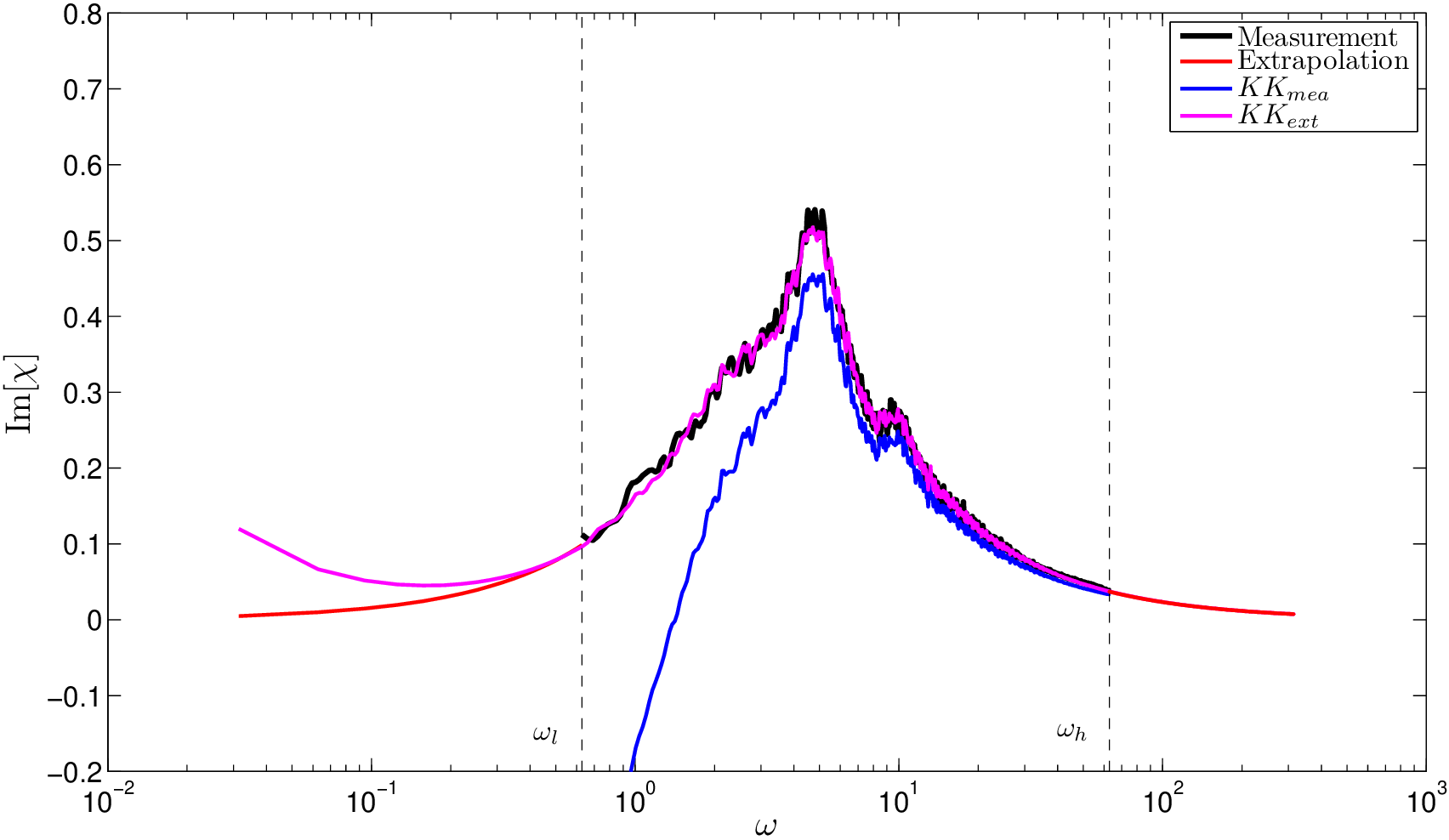}
      \end{center}
    \caption{Linear susceptibility of the energy at the grid point where the local perturbation is applied. The real and the imaginary parts are depicted in Panels a) and b), respectively. The measured and extrapolated values are shown in red and black lines, respectively. The result of the Kramers-Kronig inversion done with the measured and with with the extrapolated data are shown in blue and magenta lines, respectively.}
  \label{chienergyuno}
\end{figure}

We then proceed to analyze more in detail the linear susceptibilities related to local observables. In Fig. \eqref{chienergyuno} we present our results concerning the real and imaginary part of $\chi^{(1)}_{E_j,1}(\omega)$. Analogously to what observed in the previous subsection, we have that once the measured susceptibility is extrapolated using the theoretical results obtained via response theory and KK relations, we have an excellent agreement between the original real and imaginary parts and those obtained using the KK inversion. The KK algorithm, instead, provides only a partially satisfying outcome when only data from the measured range are considered. Relatively discrepancies are found near the boundaries of the data range, with an especially serious bias near $\omega_l$ for the imaginary part of the susceptibility.

When comparing $\chi^{(1)}_{E_j,1}(\omega)$ and $\chi^{(1)}_{E,1}(\omega)$ (see Fig. \ref{tuttiuno}), we observe that for low frequency the response of the energy at the grid point where the perturbation is applied accounts for about half of the response of the total energy, thus implying that the remaining half is redistributed among the remaining $N-1$ grid points. The relevance of the response of grid points other than the directly perturbed one also explains why the peak of Im$\chi^{(1)}_{E,1}(\omega)$ (and so of Im$\chi^{(1)}_{\varepsilon,N}(\omega)$) than that of Im$\chi^{(1)}_{E_j,1}(\omega)$ - see the frequency range $2 \lesssim \omega \lesssim 4$. Slower perturbations allow other grid points $x_k\neq x_j$ to respond effectively.

Instead, since the leading asymptotic order of $\chi^{(1)}_{E_j,1}(\omega)$ and $\chi^{(1)}_{E,1}(\omega)$  is the same, at high frequencies the local energy response accounts for most of the energy response of the whole system. In this case, the incoming perturbation is so fast that the internal time scales of the system as bypassed, and mainly a local effect is observed. Nevertheless, the second leading order of the asymptotic expansion of the two susceptibilities has opposite sign (see Eqs. \eqref{chiensingle} and \eqref{chienuno}), which suggests that at any large but finite frequency the local energy response is only a good approximation to the response of the total energy. The changeover between the two regimes occurs around the frequency of the peak of Im$\chi^{(1)}_{E_j,1}(\omega)$, which corresponds to a perturbation with period close to 1.

\begin{figure} 
   \begin{center}
   \includegraphics[width=0.5\textwidth,angle=270]{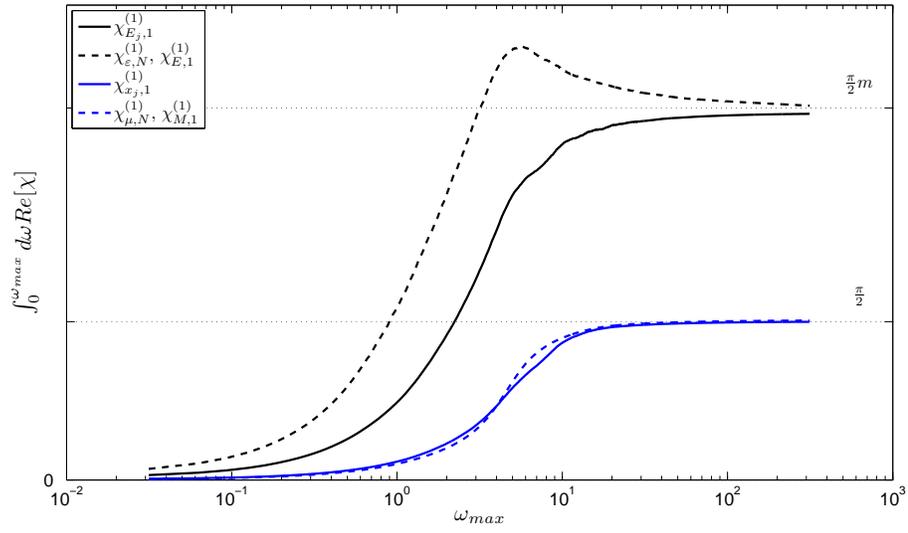}
      \end{center}
    \caption{Sum rules of the real part of the linear susceptibilities indicated in the legend. The theoretical values are indicated in the figure.}
  \label{sumrulecomparison}
\end{figure}

Thanks to the asymptotic equivalence between $\chi^{(1)}_{E_j,1}(\omega)$ and $\chi^{(1)}_{E,1}(\omega)$ (and $\chi^{(1)}_{\varepsilon,N}(\omega)$), they must obey the same sum for the real part of the susceptibility (see Eqs. \eqref{sumrulensingle}-\eqref{sumrulenoneonepert}), even if the two real parts, as discussed above, are rather different in value in the low frequency range and even in sign in the high frequency range. Figure \ref{sumrulecomparison} confirms that this rather counter-intuitive behavior is actually observed. Note also that sum rules, resulting from an integration, are less sensitive to noise in the data, but this occurs if and only if the underlying signal is correct. Therefore, we understand that in $\chi^{(1)}_{E,1}(\omega)$  and $\chi^{(1)}_{\varepsilon,N}(\omega)$ the strong static and quasi-static response and the (rather odd) negative sign for high frequencies of the real part of the linear susceptibility, which are crucially related to the behavior for the grid points different from the perturbed one) compensate each other to guarantee agreement with the sum rule obtained from the real part of $\chi^{(1)}_{E_j,1}(\omega)$, which instead has a smaller range and more regular (monotonic) behavior with frequency.

A formally similar - and analogously spectacular - spectral compensation has been observed in a physical process as different from what we are analyzing here as the electromagnetically induced transparency \cite{cata1997}. The result obtained here supports previous findings obtained on quasi-equilibrium systems suggesting that sum rules do not depend on many-particle interactions \cite{ValBass1,LSP05}.

\begin{figure} 
   \begin{center}
   \includegraphics[width=0.5\textwidth,angle=270]{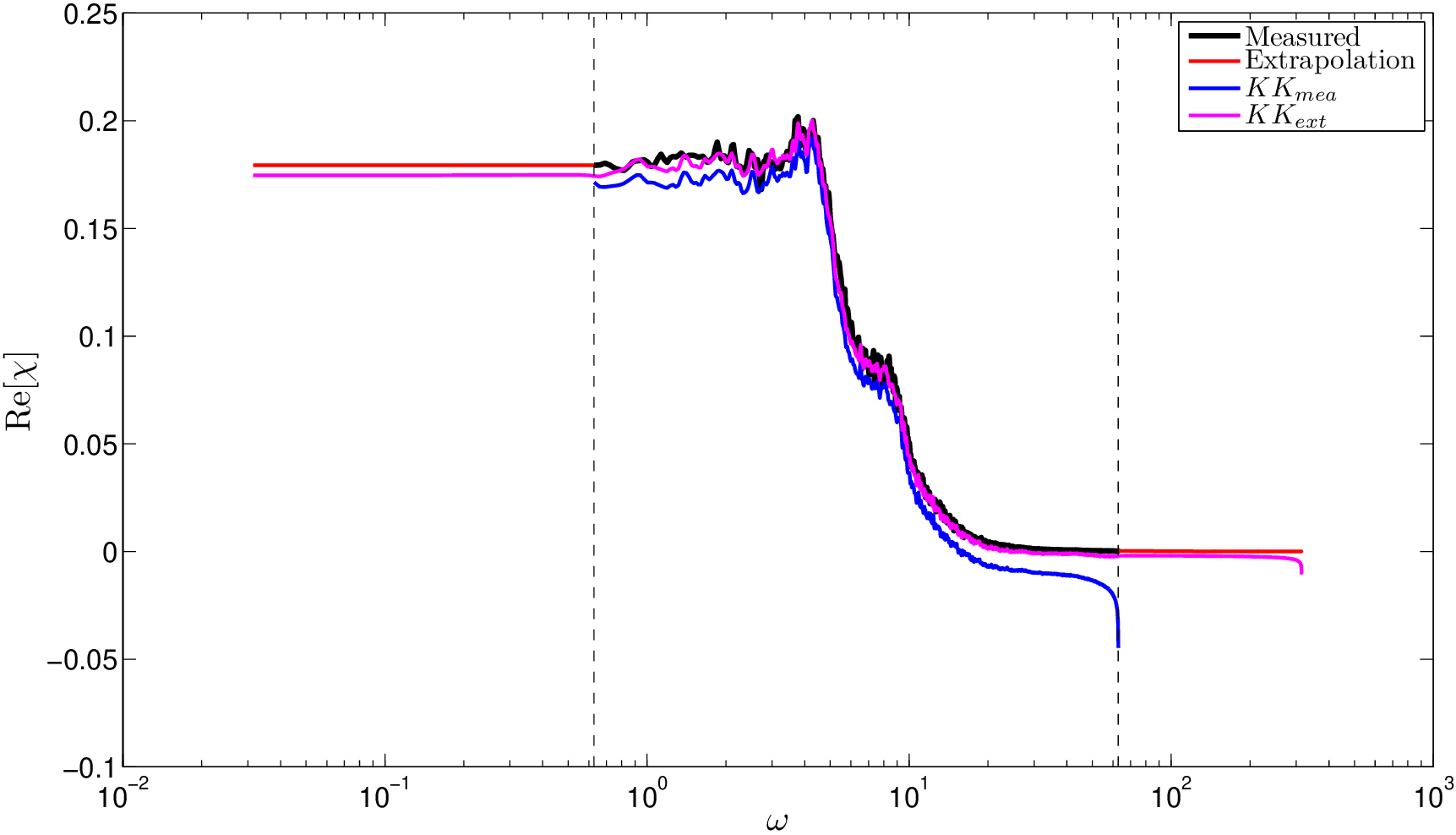}
   \includegraphics[width=0.5\textwidth,angle=270]{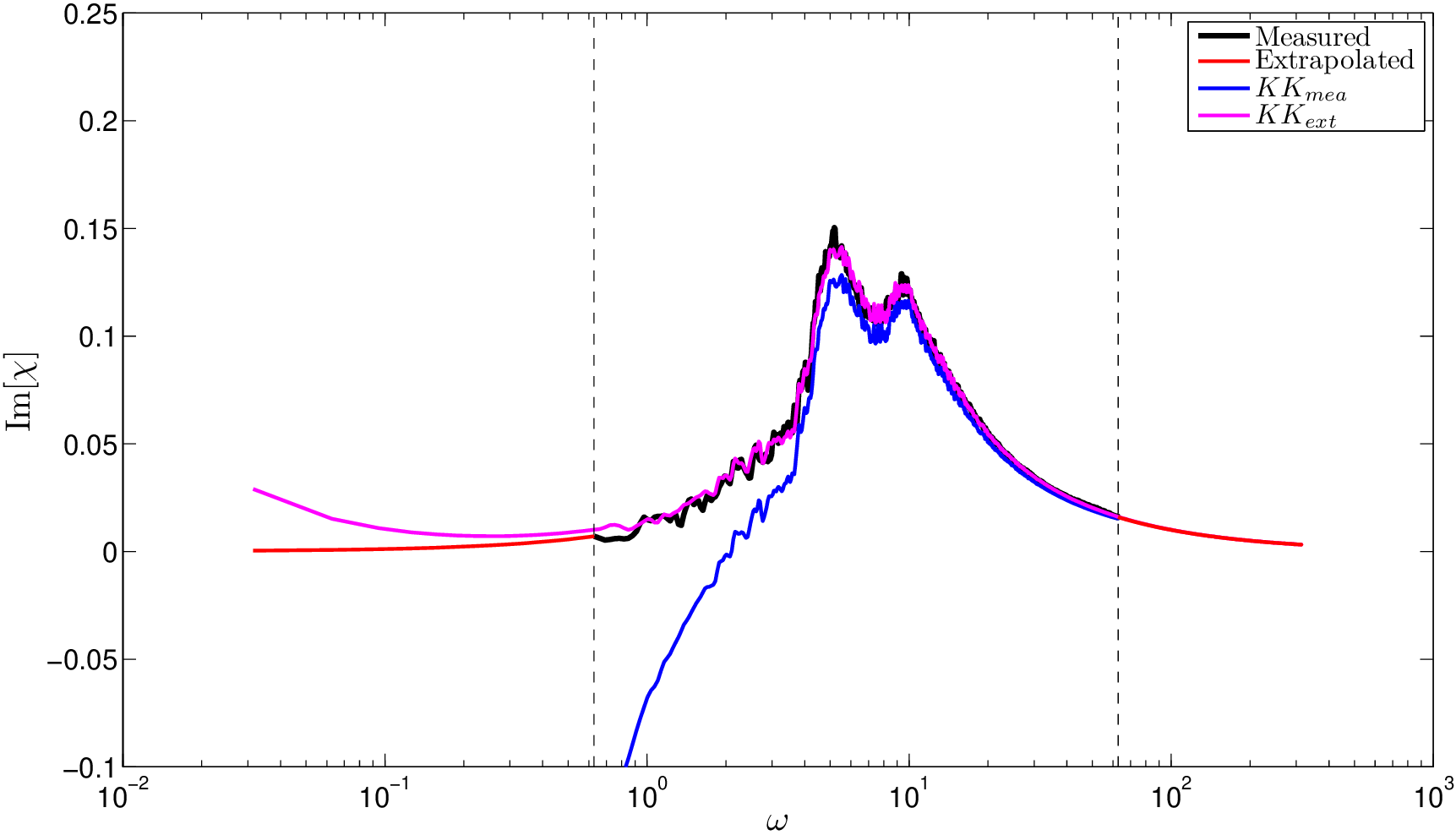}
      \end{center}
    \caption{Linear susceptibility of the momentum at the grid point where the local perturbation is applied.
    The real and the imaginary parts are depicted in Panels a) and b), respectively. The measured and extrapolated
    values are shown in red and black lines, respectively. The result of the Kramers-Kronig inversion done with the
    measured and with with the extrapolated data are shown in blue and magenta lines, respectively.}
  \label{chimomeuno}
\end{figure}

The investigation of the linear susceptibility of the variable $x_j$ is not as insightful as that of $E_j$. We find that linear susceptibility $\chi^{(1)}_{x_j,1}(\omega)$ is quite similar to $\chi^{(1)}_{M,1}(\omega)$ (and $\chi^{(1)}_{\mu,N}(\omega)$, see Fig. \ref{chimom}) in both the real and imaginary parts at all frequency. The only notable differences are that the static response $x_j$ is slightly larger than than of $M$, and that the imaginary features a secondary peak at slightly larger frequencies than the main spectral feature. We have verified,as in the previous cases, the results of the numerical simulations accurately agree with the theoretical results regarding the asymptotic behavior of both the real and imaginary part and that KK relations map to high degree of precision the real and the imaginary parts into each other. See Fig. \ref{chimomeuno} for details.

We present as main finding of the analysis of the observable $x_j$ that, as predicted by the theory, the real part of $\chi^{(1)}_{x_j,1}(\omega)$ obeys the same sum rule as the real part of $\chi^{(1)}_{M,1}(\omega)$ or of $\chi^{(1)}_{\mu,N}(\omega)$, because the corresponding imaginary parts feature the same asymptotic behavior. Figure \ref{sumrulecomparison} shows that in the case of the momentum variables the cumulative integral is rather similar for the susceptibility of the local and of the global variable, with small discrepancies in the region around the peak of the response.

\subsection{Further implications of Kramers-Kronig relations and sum rules}
We now show how the knowledge of the asymptotic behavior of the real and imaginary part and the knowledge of the validity of the KK relations and related sum rules allow to draw general conclusions on the similarities and differences between two given linear susceptibility functions. Let's consider the case that these two susceptibilities feature the same first order asymptotic expansion in the high frequency limit. Let's assume that it is an odd power of $\omega$, so that the real part is negligible for high frequencies. Therefore, the two susceptibilities will obey the same sum rule for, e.g. the $0^{th}$ moment of the real part.

If they agree also in the asymptotic behavior of the real part, they cannot feature large discrepancies in the low frequency range of the real part of the susceptibility either, or otherwise the agreement of the sum rules would be broken. Therefore, the real part of the two susceptibilities are similar, and, as a consequence of the KK relations, the two imaginary parts will also be similar.

If, instead, there is a discrepancy in the asymptotic behavior of the real part of the two susceptibilities, the two real parts will necessarily be rather different in the low frequency range, again in order to comply with the sum rule constraint. As the two real parts are different, the imaginary part of the two susceptibility will also be rather different, except, from hypothesis, in the high-frequency range.

The first scenario envisioned here pertains to the pair of linear susceptibilities $\chi^{(1)}_{x_j,1}(\omega)$ and $\chi^{(1)}_{M,1}(\omega)$, whereas the second scenario is related to the pair of linear susceptibilities $\chi^{(1)}_{E_j,1}(\omega)$ and $\chi^{(1)}_{E,1}(\omega)$. Note that, taking into account the asymptotic properties of the susceptibility of the observable $E_{loc}=1/2x_j^2+1/2x_{j+1}^2+1/2x_{j+2}^2+1/2x_{j-1}^2$, discussed in the previous section we conclude that $\chi^{(1)}_{E_{loc},1}$ and $\chi^{(1)}_{E,1}$ should be similar for all values of $\omega$.

Obviously, a similar argument applies if the leading order is real. This discussion further clarifies that the higher the number of independent KK relations and related sum rules verified by a susceptibility functions, the more stringent are the constraints on its properties.

\subsection{Additional Properties of the Linear Susceptibility}
\begin{figure} 
   \begin{center}
   \includegraphics[width=0.5\textwidth,angle=270]{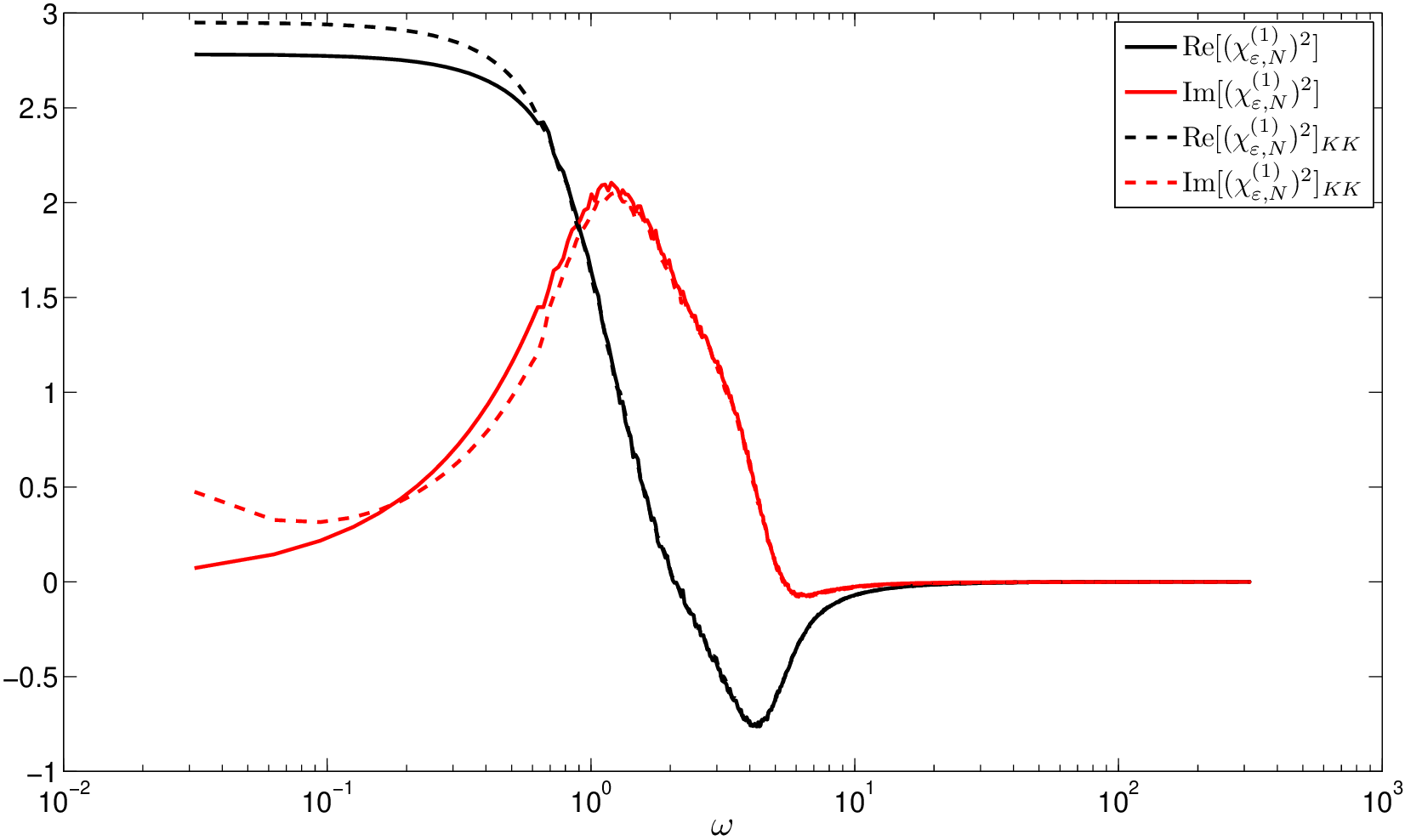}
      \includegraphics[width=0.5\textwidth,angle=270]{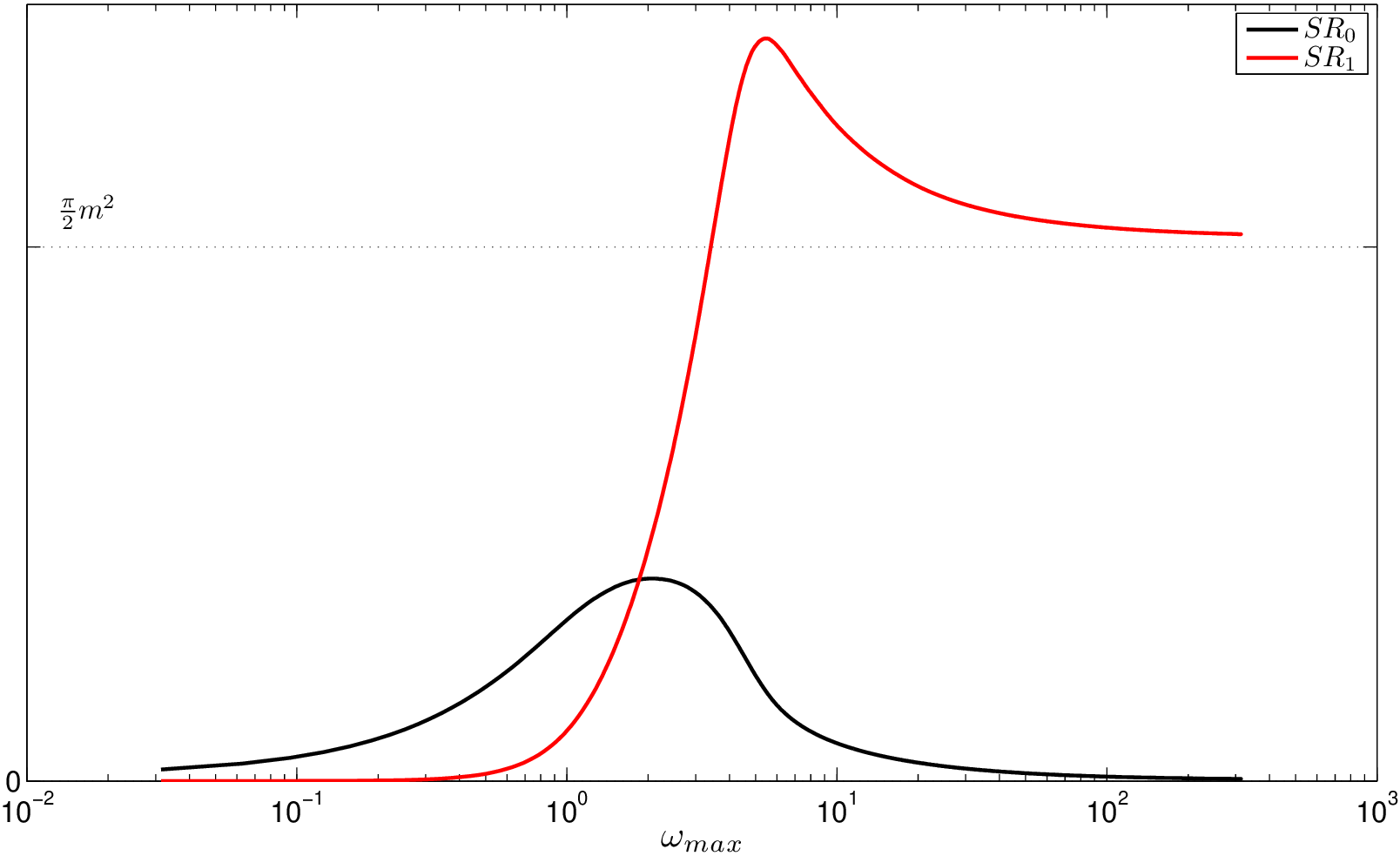}
      \end{center}
    \caption{Properties of the square of the linear susceptibility $\chi^{(1)}_{\varepsilon,N}$. The real and imaginary parts of $[\chi^{(1)}_{\varepsilon,N}]^2$ with their KK transforms are depicted in the first panel, the vanishing sum rule for the real part and the non-vanishing sum rule for the imaginary part are depicted in the second panel.}
  \label{chisquare}
\end{figure}

The special mathematical properties of the linear susceptibilities allow to investigate further properties of the response. In particular, we note that for $m\geq 1$ the function $[\chi_\Phi^{(1)}]^m$ is analytic in the upper complex $\omega$-plane just as as $\chi_\Phi^{(1)}$. This allows, as discussed in \cite{LSP05} to derive, in principle, an infinite set of integral relations (KK and sum rules) deriving just from the holomorphic proprieties of the susceptibility. As an example, we have considered the square of the linear susceptibility $[\chi_{\varepsilon,N}^{(1)}(\omega)]^2$. From Eq. \eqref{chiensingle}, it is easy to prove that the following asymptotic expansion holds for large values of $\omega$:
\begin{equation}
[\chi_{\varepsilon,N}^{(1)}(\omega)]^2=-\frac{m^2}{\omega^2}+\frac{m(F-2m)}{\omega^3}+o(\omega^{-4}).
\label{asymsquare}
\end{equation}
As shown in the first panel of Fig. \ref{chisquare}, KK relations are found to connect up to a high degree of approximation the real and imaginary part of $[\chi_{\varepsilon,N}^{(1)}(\omega)]^2$. Moreover, thanks to the asymptotic behavior given in Eq. \eqref{asymsquare}, it is possible to establish, thanks to Eqs. \eqref{vanishsum2}-\eqref{evensumrule}, the following sum rules:
\begin{equation}
\int^{\infty}_{0} \textrm{d}\nu\textrm{Re}[\chi_{\varepsilon,N}^{(1)}(\nu)]^2 \,  = 0 ,
\end{equation}
\begin{equation}
\int^{\infty}_{0} \textrm{d}\nu \nu \textrm{Im}[\chi_{\varepsilon,N}^{(1)}(\nu)]^2\,  =\frac{\pi}{2} m^2,
\end{equation}
The second panel of Fig. \ref{chisquare} shows that the obtained numerical results are in excellent agreement with the theoretical predictions. Note that these results do not have an obvious physical interpretation, as the inverse Fourier Transform of $[\chi_{\varepsilon,N}^{(1)}(\omega)]^2$ is given by the convolution product of the Green function $G_{\varepsilon,N}^{(1)}(t)$ with itself, while they depend only on the formal properties of the linear susceptibility.

\section{Practical Implications for Climate Change Studies}
\label{change}
In this paper we have constructed and verified to a high degree of accuracy the linear response theory for a simple yet prototypical climate model by computing the frequency-dependent susceptibilities of several relevant observables related to localized and global patterns of forcings. These results pave the way for devising a rigorous methodology to be used by climate models of any degree of complexity for studying climate change at, in principle, all time scales using only a very limited set of experiments, and for exploiting effectively the currently adopted ensemble runs methods. 

Let's consider, for sake of simplicity, that the observable $\Phi$ is the time-dependent globally averaged surface temperature of the planet $T_S$, that $F(x)$ represents the whole set of climate equations in a baseline scenario (e.g., with pre-industrial $CO_2$ concentration), and that the perturbation field $X(x)$ is nothing but a constant field of $CO_2$ concentration, which directly impacts only the radiative part of the code. The perturbation is modulated by a time-dependent function $f(t)$ to be specified below. We assume, for simplicity, that the model does not feature daily or seasonal variations in the radiative input at the top of the atmosphere.

From Eq. \eqref{risposta}, we have that
\begin{equation}
\left\langle T_S \right\rangle^{(1)}(t)=
 \int^{+\infty}_{-\infty}\textrm{d}\sigma_1 G_{T_S}^{(1)}(\sigma_1)f(t-\sigma_1).
\label{rispostabis}
\end{equation}
In practical terms, the left hand side of this equation is nothing but the ensemble average of the time series of the change between the globally averaged surface temperature of the planet at a time $t$ after the perturbation has started. Note that the direct estimate of $G_{T_S}^{(1)}(\sigma)$ is likely to be overwhelmingly difficult. Using Eq. \eqref{Susciforce}, we have that:
\begin{equation}
\label{Susciforcebis}
\left\langle T_S \right\rangle^{(1)}(\omega)=  \chi^{(1)}_{T_S}(\omega) f(\omega),
\end{equation}
which implies that once we compute the Fourier Transform of the time series mentioned above and we know the modulating function $f(t)$ (and so its Fourier Transform $f(\omega)$), we can reconstruct $\chi^{(1)}_{T_S}(\omega)$. Let's select a particularly simple example of modulating function $f(t)=\epsilon(\Theta(t)-\Theta(t-\tau))$. This is just a rectangular function of width $\tau$, of height $\epsilon$, and shifted from the origin by a forward time translation $\tau/2$. In practical terms, this corresponds to changing abruptly the field $CO_2$ concentration by $\epsilon$ at time $t=0$ and taking it back to its original value at $t=\tau$. we then obtain:
\begin{equation}
\label{Susciforceter}
\chi^{(1)}_{T_S}(\omega)=\frac{\left\langle T_S \right\rangle^{(1)}(\omega)}{f(\omega)}=\omega\frac{\left\langle T_S \right\rangle^{(1)}(\omega)}{\epsilon(\sin(\omega\tau)+i(1-\cos(\omega\tau)))}.
\end{equation}
Once we know $\chi^{(1)}_{T_S}(\omega)$, as widely discussed in this paper, we can compute $G^{(1)}_{T_S}(t)$, and we know everything about the response of the system at all time scales, including the static response. Note that any choice of $f(t)$ is equally valid to set up this procedure as long as $f(t)$ is square integrable. This implies that, in a very profound way, the kind of forcing scenarios used in the various assessment Reports of IPCC, where the $CO_2$ concentration typically stabilizes at a different value from the preindustrial one (so that $f(t)$ does not tend to 0 as $t\rightarrow \infty$) are not necessarily the only nor the best ones, in spite of what could be intuitively guessed, to study even the steady state response of the system.

Obviously, a similar set of experiments could be devised for studying rather thoroughly the response of the climate system to a variety of forcings, such as changes in the $O_3$ concentration, aerosols, solar radiance, as well as to changes in the parameterizations. In the case of uncoupled models of one subdomain of the climate system (e.g. atmospheric and oceanic GCMs, land-surface models), this strategy could be used to study the impact of perturbations to the boundary conditions provided by the other subdomains of the climate system.

\section{Summary, Discussion and Conclusions}
\label{conclu}
The climate can be seen as a complex, non-equilibrium system, which generates entropy by irreversible processes, transforms moist static energy into mechanical energy \cite{Lorenz67,Peix} as if it were a heat engine \cite{Johnson,Luc09}, and, when the external and internal parameters have fixed values, achieves a steady state by balancing the input and output of energy and entropy with the surrounding environment \cite{Ozawa,Luc09}. For such basic reasons, the tool of equilibrium and quasi-equilibrium statistical mechanics cannot provide suitable tools for studying the fundamental properties of the climate system. In particular, the fluctuation-dissipation theorem, which allows for deriving the properties of the response of the system to external perturbations from the observations of its internal variability cannot be applied.

It is reasonable to ask whether is possible to evaluate how far from equilibrium the climate system actually is. It is possible to evaluate such \textit{distance} in a  mathematically sound way by assessing the ratio of the dimensionality of the attractor of the system over the total number of degrees of freedom. Whereas a ratio close to one indicates that only small deviations  from equilibrium are present, a small ratio suggests that strongly non-equilibrium conditions are established. See \cite{Posch} for a detailed treatment of this problem in the classical case of heat conduction. In the case of a quasi-geostrophic atmospheric model forced by Earth-like boundary conditions, the dimensionality of the attractor of the model is about one order of magnitude smaller than the total number of degrees of freedom \cite{Vann97}. While not conclusive, this seems to suggest that the best framework to interpret the climate is that of a far from equilibrium system.

Following either explicitly or implicitly the programme of the Catastrophe theory \cite{Arnold}, many authors have approached the problem of understanding the fundamental properties of the climate system by looking at the detailed structure of the bifurcations of the deterministic dynamical system constructed heuristically in order to  represent the dynamics of the main climate modes using as few degrees of freedom as possible. Such an approach often hardly allows to efficiently represent the fluctuations and the statistical properties of the system. The introduction of stochastic forcing provides a relatively simple but conceptually rich partial solution to some of these draw-backs, even if the \textit{Hasselmann programme} \cite{Hassel} suffers from the need for a - usually beyond reach - \textit{closure theory} for the properties of noise. Therefore, the stochastic component is usually introduced ad hoc, with the ensuing lack of universality and/or robustness when various levels of truncations are considered. These strategies have anyway brought to outstanding scientific results and has been suggested the existence of generic mathematical structures present in hierarchies of CMs \cite{Saltzman}. Recently, the unified treatment of chaotic and stochastic dynamics using the results of the mathematical theory of random dynamical systems is emerging as new, promising paradigm for the investigation of the structural properties of the climate system \cite{Chekroum}.

We have proposed a different perspective. In agreement with the  view given above, we have taken as mathematical framework for the analysis of the climate system that of non-equilibrium statistical mechanics, and have focused on the steady state properties of ergodic dynamical systems \cite{EckRuelle} possessing the special property of having an invariant measure of the SRB type \cite{Ruelle2}. As proposed by the chaotic hypothesis \cite{Gallav,GallavCohen}, this mathematical framework is well suited for analyzing general non-equilibrium physical systems.

In this context, the impact on the system of general perturbation can be treated using the response theory recently introduced by Ruelle \cite{Ruelle1,Ruelle1b,Ruelle3}, which allows to compute the change in the expectation value of a generic observable as a perturbative series where each term is given by the average over the unperturbed invariant measure of a function of the phase space which depends on the considered observable and on the applied perturbation. In other terms, even if the internal dynamics of the system is nonlinear and chaotic, the leading order of the response is in general linear with the strength of the added perturbation. This approach overcomes the difficulties related to the singularity of the invariant measure discussed in \cite{Thuburn}.

At each order, the propagator of the perturbation, i.e. the Green function, is causal. This allows for applying dispersion theory and establish general integral constraints - KK relations - connecting the real and imaginary parts of the susceptibility, i.e. the Fourier Transform of the Green function \cite{(Val08),(Val09)}

In this paper we have first recapitulated the main aspects of the general response theory and have propose some new general results, which boil down to consistency relations between the linear susceptibilities of different observables. The obtained equation provides the basic idea why the fluctuation-dissipation theorem does not apply in non-equilibrium cases.

We have showed for the first time that the Ruelle linear response theory can be applied with great success to analyze the climatic response to general forcings. We have chosen as test bed the Lorenz 96 model \cite{Lorenz3,LorenzEmm,Lorenz2}, which, in spite of its simplicity, has a well-recognized prototypical value as it is a spatially extended one-dimensional model and presents the basic ingredients, such as dissipation, advection and the presence of an external forcing, of the actual atmosphere. Such a model features a different level of complexity with respect to those adopted in previous numerical investigations of Ruelle's theory \cite{Reick,CessSep,(Val09)}

We have analyzed the frequency dependence of the response of the local and global energy and momentum of the system to perturbations having a global spatial pattern and to perturbations acting only on one grid point. We have derived analytically several properties of the corresponding susceptibilities, such as asymptotic behavior, validity of KK relations, and sum rules. We have shown that all the coefficients of the leading asymptotic expansions as well as the integral constraints can be written as linear functions of parameters that describe unperturbed properties of the system, and in particular its average energy and average momentum. The theory has been used to explain differences in the response of local and global observables, in defining the intensive properties of the system and in generalizing the concept of climate sensitivity to all time scales.

We have then verified the theoretical predictions from the outputs of the simulations up to a high degree of precision, even if we have used rather modest computational resources (a total of about 30 cpu days of a mid-range commercial laptop). We have verified that the linear response theory holds for perturbations of intensity accounting to up to about 10\% of the unperturbed forcing terms. Even when local perturbation and local observables are considered it is possible to achieve a signal-to-noise ratio which permits rather satisfactory comparisons with the theory. We have proved that the combined use of KK relations and the knowledge of the asymptotic behavior of the susceptibilities allows for extrapolating in a rigorous way the observed data. We also have shown how to reconstruct the linear Green function, which can be used to map perturbations of general time modulation into changes in the expectation value of the considered observable for finite as well as infinite time.

Our numerical experiments have been performed using one of the standard settings of the Lorenz 96 model, namely the version identified by having $N=40$ degrees of freedom and forcing $F=8$. Nevertheless, some newly obtained empirical closure equations expressing the average energy and the average momentum of the unperturbed system as simple power laws of $F$ (with no dependence on $N$) have allowed to extend our results to the entire class of chaotic Lorenz 96 models.

In this paper we have only used the KK relations in the most simplistic framework, i.e., computing the KK transforms and evaluating their agreement with the original data. Actually, several more sophisticated analysis techniques are available, such as recursive self-consistent algorithms, where the measured data are taken as first guess, exploiting the fact that multiple applications of KK relations, combined with sum rules, automatically filter our the noise and remove most of the spurious signal \cite{LSP05}.

We believe that the proposed approach, which we may dub as \textit{spectroscopy of the climate system}, may constitute a mathematically rigorous and practically very effective way to approach the problem of evaluating climate sensitivity and climate change from a radically new perspective. In this regard, we have proposed a rigorous way to compute the surface temperature response to changes in the the $CO_2$ concentration at all time scales using only a specific set of simulations, and taking advantage of the theoretical results presented here. We underline that our approach takes into account all the (linear and nonlinear) feedbacks of the system, as they are included in the definition of the Green function. This, at a very practical level, is the great advantage of using Ruelle's formulas.

At a more basic level, whereas considering more complex models requires heavier computational resources, the modest cost of the present set of simulations suggests that, at least for global or regional climatic observables, it is feasible to test the theory discussed here for simplified yet Earth-like climate models without resorting to top-notch computing facilities. Moreover, while in this paper we have computed the susceptibilities using, on purpose, a very cumbersome method, more efficient strategies can be devised, at least when the linear regime of the response is considered. Apart from the practical example given for the case of the impact of the $CO_2$ concentration, these include studying the response of the system to $\delta(t)$ like perturbations, which gives directly the Green function of the system, and including in the forcing various monochromatic signals. Of course, in all cases, a Monte Carlo approach is needed in order to sample effectively the attractor of the unperturbed system in terms of the initial conditions of the simulations.

These results pave the way for future investigations aimed at improving and extending the theoretical framework presented here, at finding results of general applicability in the context of the modelling of geophysical fluid dynamics, and, finally at answering specific questions of relevance for climate dynamics. In this paper we have analyzed the simple case of the linear response, but, as discussed in \cite{Ruelle1b,(Val08),(Val09)}, we have the algorithm to compute higher order terms, so that the treatment of the nonlinear response in entirely feasible.

In the first direction, we foresee the possibility of writing out explicitly the linear susceptibility of a general observable by projecting the perturbation onto the unstable, neutral and stable manifolds and analyzing separately the contributions to the total response. This will probably require the adoption of adjoint techniques. Moreover, we will be testing the radius of convergence of the Ruelle response theory is some specific examples.

Along the second direction, we propose to study the impact of stochastic forcing to deterministic chaotic models by treating the (additive or multiplicative) noise as a perturbation to be analyzed using the linear and nonlinear Ruelle response theory and related spectral methods. Moreover, we shall look into the spectral peaks of the susceptibilities and try to understand how the amplification of the response is related to resonances of the system and to the activation of positive feedbacks.

Along the third direction, we envision the analysis of the impact of topography on the statistical properties of the circulation in a quasi-geostrophic setting, thus extending in a climatic perspective what presented in \cite{Spe85}. Moreover, we will tackle in an idealized setting the problem of computing the response of the storm track to changes in the surface temperature \cite{Brayshaw}. Moreover, we will try to compute along the way discussed here the climate response to changes in $CO_2$ and solar irradiance using simplified but rather valuable climate models like PLASIM \cite{Frae05}.

Finally, we would like to remark that the theory and the practical recipes proposed here could be of direct interest for all projects aimed at auditing climate models' performances and at studying practical problems connected to climate change, such as PCMDI/CMIP3 (\texttt{http://www-pcmdi.llnl.gov/}), distributed computing initiatives such as \texttt{climateprediction.net}, and the new project PCMDI/CMIP5 (\texttt{http://cmip-\\pcmdi.llnl.gov/cmip5/}), which will provide a crucial input for the Fifth Assessment Report of the IPCC.
\\
\\
\thanks{\textit{VL acknowledges the financial support of the EU-ERC research grant NAMASTE and of the Walker Institute Research Development Fund.}}

\newpage
\clearpage

\end{document}